\documentclass[usenatbib]{mn2e}
\usepackage{natbib}

\usepackage{amsmath,url}
\usepackage{latexsym,graphicx,natbib,amssymb}
\usepackage{units}
\usepackage{color}

\usepackage{amsfonts}% AMS Fonts
\usepackage{textcomp}% einige Textsymbole (°C usw.)
\usepackage{verbatim}%Quellcode anzeigen
\usepackage{psfrag}%Labels im ps/eps Figures nachtraeglich editieren
\usepackage{epsfig}
\usepackage{caption}
\usepackage{float}

\newcommand{\aj}{AJ}
\newcommand{\apj}{ApJ}
\newcommand{\apjl}{ApJ}
\newcommand{\apjs}{ApJ}
\newcommand{\mnras}{MNRAS}
\newcommand{\pasp}{PASP}
\newcommand{\pasj}{PASJ}
\newcommand{\araa}{A\&AR}
\newcommand{\aap}{A\&A}

\newcommand{\nat}{Nat}

\newcommand{\sgra}{Sgr~A$^\star$}
\newcommand{\mbh}{M_{\bullet}}
\newcommand{\msun}{{\rm M}_{\odot}}

\newcommand\simless{\mathbin{\lower 3pt\hbox
   {$\rlap{\raise 5pt\hbox{$\char'074$}}\mathchar"7218$}}}
\newcommand\simgreat{\mathbin{\lower 3pt\hbox
   {$\rlap{\raise 5pt\hbox{$\char'076$}}\mathchar"7218$}}}

\newcommand{\mstar}{M_{\star}}

\newcommand{\rstar}{r_{\star}}

\newcommand{\rsun}{{\rm R}_{\odot}}

\newcommand{\jmbh}{J_{\bullet}}

\usepackage{caption}

\title[Massive Black Holes in Galaxies]
      {Tidal disruption rate of stars by supermassive black holes obtained by direct N-body
simulations}
\date{Received / Accepted}

\pagerange{\pageref{firstpage}--\pageref{lastpage}} \pubyear{2011}

\author[M. Brockamp et  al.]
{M.~Brockamp$^1$\thanks{brockamp@astro.uni-bonn.de}\thanks{Member of the International Max Planck Research School (IMPRS) for Astronomy and Astrophysics at the Universities of Bonn and Colone}, H.~Baumgardt$^2$\thanks{h.baumgardt@uq.edu.au}, P.~Kroupa$^1$\\
$^1$ Argelander Institute for Astronomy (AIfA), Auf dem H\"ugel 71, D-53121 Bonn, Germany \\
$^2$School of Mathematics and Physics, University of Queensland, Brisbane, QLD 4072, Australia \\ }
\begin{document}
%\pagerange{\pageref{firstpage}--\pageref{lastpage}} %\pubyear{2009}

\maketitle

\label{firstpage}

\begin{abstract}
The disruption rate of stars by supermassive black holes (SMBHs) is calculated numerically with a modified version of Aarseth's NBODY6 code. Equal-mass systems without primordial binaries are treated.
The initial stellar distribution around the SMBH follows a S\'{e}rsic $n=4$ profile representing bulges of late
type galaxies as well of early type galaxies without central light deficits, i.e. without cores.
In order to infer relaxation driven effects and to increase the statistical significance, a
very large set of N-body integrations with different particle numbers N, ranging from $10^{3}$ to $0.5\cdot10^{6}$ particles, is performed. 
Three different black hole capture radii are taken into account, enabling us 
to scale these results to a broad range of astrophysical systems with relaxation times shorter than one Hubble time, i.e. for SMBHs up to $\mbh \approx \unit[10^{7}]{\msun}$.
The computed number of disrupted stars are driven by diffusion in angular momentum space into the loss cone of the black hole and the rate scales with the total number of particles as $\frac{\text{d}N}{\text{d}t}\propto N^{b}$,
where $b$ is as large as 0.83. This is significantly steeper than the expected scaling
$\frac{\text{d}N}{\text{d}t}\propto \ln(N)$ derived from simplest energy relaxation arguments. 
Only a relatively modest dependence of the tidal disruption rate on the mass of the SMBH is found and we
discuss our results in the context of the $\mbh-\sigma$ relation.
The number of disrupted stars contribute a significant part to the mass growth of black holes in the lower mass range as long as a significant part of the stellar mass becomes swallowed by the SMBH.  
This also bears direct consequences for the search and existence of IMBHs in globular clusters. 
For SMBHs similar to the galactic center black hole \sgra, a tidal disruption rate of $55\pm27$ events per Myr is deduced.
Finally relaxation driven stellar feeding can not account for the masses of 
massive black holes $\mbh \geq \unit[10^{7}]{\msun}$ in complete agreement with conventional gas accretion and feedback models.

\end{abstract}


\section{Introduction}
The evolution of supermassive black holes (SMBHs) and their host galaxies is at present one of the key problems of astrophysics.
Motivated by empirically found scaling relations between properties of galaxies in terms of velocity dispersion
$\sigma$ \citep{Gebhardt2000, Ferrarese2000, Richstone2009}, luminosity $L$ \citep{Kormendy1995, Ford2005}, bulge mass $M_{\text{Bulge}}$ \citep{Magorrian1998, Rix2004},
central light deficit $L_{\text{def}}$ \citep{Lauer2007,Kormendy2009, Hopkins2010}, total number of globular clusters $N_{\text{GC}}$ \citep{Burkert2010} and the mass of their central black holes $\mbh$, there is
a substantial need to understand the related evolution of both SMBHs and their hosts.
In order to constrain galaxy formation models and to answer the question as to what powers the growth of SMBHs over cosmic times, all forms of matter
which are accreted must be taken into account.
This becomes more urgent as recent studies have found evidence for deviations from the general scaling relations for the most-massive and for the least-massive black holes \citep{Lauer2007, Gebhardt2011, Kormendy2011}.\\
  
Gas accretion is thought to be the most dominant driver of SMBH growth \citep{Soltan1982}. Modern studies \citep{Yu2002} estimate the black hole mass density from 
the spatial distribution and from the measured stellar velocity dispersions in elliptical galaxies in combination with the $\mbh-\sigma$ relation. The SMBH mass density is then compared with the observed 
quasar luminosity function in order to yield constraints on the accretion efficiency parameter $\epsilon$ as well as on the growth history.
In order to make these studies even 
more accurate, the impact of other feeding modes like 
merging supermassive black holes and stellar captures must also be taken into account.
Simultaneously the luminous gas accretion
history of low-mass SMBHs ($\mbh\approx \unit[10^{5}-10^{7}]{\msun}$) is 
harder to measure especially at large redshifts as they never approach luminosities comparable to those of quasars. It is even plausible that low-mass SMBHs gain
most of their mass by tidal disruption events \citep{Ho2006}. 
Therefore it is important to infer the stellar capture rate for as many astrophysical systems of interest as possible, for all relevant SMBH masses using both theoretical
and when possible numerical approaches.
In order to avoid confusion regarding the terminology of the capture and disruption rate we note that the former expression is used for  
the general number of stars/particles which are either swallowed as a whole or disrupted outside the event horizon in a given time, i.e. independent of the mass of the SMBH. The latter one is explicitly used for
situations in which stars are tidally disrupted before they would enter the event horizon. \\

In this paper we present the disruption rate of stars by SMBHs with masses  
in the lower range up to $\mbh \lesssim \unit[10^{7}]{\msun}$ embedded inside realistic stellar density profiles.
These results are obtained by self-consistent direct N-body integrations and increase the hitherto probed region of direct numerically inferred disruption
rates. Pioneered by \cite{Baumgardt2004a, Baumgardt2004b, Baumgardt2006} for intermediate-mass black holes (IMBHs) at the centers of globular clusters, our calculations can be applied to a 
larger sample of systems. 
Our findings should be regarded as complementary to other contributions \citep{Duncan1983, Magorrian1999, Amaro-Seoane2004} where the impact
of tidal disruption events is shown to be significant and therefore should not be 
neglected in considering the question of what powers the growth of black holes.\\ 

There are several mechanism by which stars are driven into the loss cone of a black hole.
In spherical stellar distributions, where the relaxation time $T_{\text{rel}}$ is comparable to or 
smaller than the present age of the universe\footnote{For the purposes of this study we do not discriminate between $t_{0}$ and one Hubble time $H_{0}^{-1}$ and
assume $H_{0}^{-1}\approx t_{0}=\unit[13.7\cdot10^{9}]{yr}$ \citep{Komatsu2009}.} $t_{0}$ \citep{Freitag2008}, two-body relaxation induces a steady change in the angular momentum space distribution of stars
such that some of them will drift to very eccentric orbits with pericentre distances smaller than the black hole capture radius \citep{Rees1976, Lightman1977}. 
In much larger systems like the most-massive elliptical galaxies which are thought to be triaxial in shape \citep{Kormendy1996}, stars on box orbits can cross
the central region arbitrary close to the SMBH \citep{Binney2008} such that they become disrupted or swallowed as a whole for the case of a very massive SMBH. 
\cite{Merritt2010} concluded that the feeding mode of very massive SMBHs, like M87 \citep{Gebhardt2009}, 
is currently dominated by stellar captures. The true rates could be even higher since their analysis takes only stellar orbits within the black hole influence radius $r_{h}$ into account,
whereas stars within the loss cone but from much further away should reach the black hole, too, as long as the critical radius $r_{\text{crit}}$ (a quantity which is defined in Eq.~\ref{F15}) remains larger than
$r_{h}$. \cite{Norman1983, Poon2001, Poon2002, Merritt2004, Berczik2006} provide additional information on the dynamics of SMBHs and stellar
capture rates in triaxial potentials.\\

Observed disruption events (\citealt{Ulmer1999, Komossa2002, Halpern2004, Komossa2004, Esquej2008, Gezari2008, Cappellut2009, Gezari2009, Komossa2009, vanVelzen2009} and references therein), support the
view that tidal disruptions contribute to the growth history of SMBHs. To which magnitude this is the case is a major aspect of this study.\\
 
The paper is organized as follows. In \S~\ref{2} we will shortly explain the concept by which stars are driven by angular momentum diffusion into the ``loss cone'' of 
the SMBH. This formalism is applied to spherical stellar distributions with arbitrary slope parameters of the density profile. \S~\ref{NBODY} describes the NBODY6 code that we used. We will specify the scale-free 
models and motivate the very large set of performed simulations required to infer the disruption rate of stars by SMBHs in the nuclei of galaxies. The
results will be given in \S~\ref{5} while more detailed information regarding the dynamics of the simulations will be part of \S~\ref{3}.
In \S~\ref{6} the procedure how to scale the obtained results to realistic astrophysical systems as well as the number of expected tidal disruption events will be specified.  
A critical discussion of potential error sources in \S~\ref{5.3} is followed by a summary of our main findings in \S~\ref{7}.

\section{Theory}\label{2}
\cite{Rees1976} calculate that massive black holes can grow not only by swallowing stars which lose their energy via
dynamical relaxation, but also by swallowing stars on very eccentric orbits i.e.\ stars with low angular momentum. The change of the stellar distribution in angular momentum space is expected to progress much faster
than the change in energy space for stars within the critical radius $r_{\text{crit}}$. 
Considering stars with very low angular momentum and pericentre distances smaller than the capture radius of the black hole, the velocity vectors of these stars
must be aligned very narrowly. This narrow region is known as the loss cone.
It only reflects the geometry in velocity space and is characterized by the loss cone angle 
$\theta_{\text{lc}}$ whose symmetry axis is directed towards the position of the black hole. For distances below the influence radius $r_{H}$ of the black hole where
the velocity profile follows a Keplarian one ($\propto r^{-0.5}$), $\theta_{\text{lc}}$ is given by
\begin{equation}\label{F13}
 \theta_{\text{lc}}\propto \left(\frac{2r_{\text{cap}}}{3r}\right)^{\frac{1}{2}} 
\end{equation}   
according to \cite{Rees1976}. For $r \geq r_{H}$ a slightly different expression has to be used. At the moment we leave
it undefined if the stars are disrupted before entering the horizon of the black hole or if they are swallowed as a whole. 
A general capture radius $r_{\text{cap}}$ can be specified for all purposes (\citealt{Novikov1989, Binney2008}; see also Appendix~\ref{AppendixA}). 
In perfectly spherical potentials i.e.\ potentials where no torques from anisotropic matter distributions
can induce an additional supply of stars, all stars on loss cone orbits would be consumed within one orbital time scale $t_{\text{cross}}$.
However, dynamical relaxation between stars causes a steady change of the stellar distribution in angular momentum space and therefore changes in the velocity vectors by small amounts $ \theta_{\text{Diff}}$ per 
crossing time \citep{Rees1976}:
 \begin{equation}\label{F14}
 \theta_{\text{Diff}}\propto \left(\frac{t_{\text{cross}}}{t_{\text{rel}}}\right)^{\frac{1}{2}}. 
\end{equation}  
The critical radius $r_{\text{crit}}$ which is the characteristic distance to the black hole where the drift in the velocity vector of a star due to dynamical relaxation within one crossing
time is of the same order as $\theta_{\text{lc}}$ is therefore defined as:
\begin{equation}\label{F15}
\frac{\theta_{\text{lc}}}{\theta_{\text{Diff}}}\Bigg|_{r=r_{\text{crit}}}=1
\end{equation}
Assuming a number density profile\footnote{The parameter $n_{0}$ can be substituted by $n_{0}=n_{c}r_{H}^{-\alpha}$ into the more common number density $n_{c}$ at the influence radius $r_{H}$.}
$n(r)=n_{0}r^{\alpha}$ within $r_{\text{crit}} \le r_{H}$ and considering only equal mass stars, an expression for the critical radius  
\begin{equation}\label{F16}
r_{\text{crit}}\propto \left(\frac{r_{\text{cap}}\mbh^{2}}{\mstar^{2}n_{0}} \right )^{\frac{1}{4+\alpha}}
\end{equation}
is obtained by inserting Eq.~\ref{F13} and Eq.~\ref{F14} into Eq.~\ref{F15}.
Spitzer's relaxation formula \citep{Spitzer1958, Spitzer1987} is used for the relaxation time $t_{\text{rel}}$. The Coulomb logarithm is neglected.\\

The stellar capture rate can be derived by using eq.~17 from \cite{Rees1976}\footnote{We replace $v(r)\propto\frac{r}{t_{\text{cross}}}$.}:  
\begin{equation}\label{F17}
\dot{C} \propto \frac{\theta_{\text{lc}}^{2}r^{3}n(r)}{t_{\text{cross}}}\Bigg|_{r=r_{\text{crit}}}=\frac{\theta_{\text{Diff}}^{2}r^{3}n(r)}{t_{\text{cross}}}\Bigg|_{r=r_{\text{crit}}},
\end{equation}
For a density profile $n(r\leq r_{\text{crit}})=n_{0}r^{\alpha}$ the stellar disruption rate $\dot{C}$ is obtained by 
substituting $r=r_{\text{crit}}$:
\begin{equation}\label{F18}
\dot{C} \propto G^{\frac{1}{2}}\mbh^{\frac{1}{2}} r_{\text{cap}} n_{0} \left(\frac{r_{\text{cap}} \mbh^{2}}{\mstar^{2}n_{0}} \right)^{\frac{0.5+\alpha}{4+\alpha}}.  
\end{equation}
For very massive black holes the critical radius
becomes larger than the influence radius of the black hole and Eq.~\ref{F16} must be modified according to \cite{Rees1976}:
\begin{equation}\label{F19}
r_{\text{crit}}\propto \left(r_{\text{cap}}r_{H}n_{0}\right )^{-\frac{1}{1+\alpha}}.
\end{equation}
We assume the velocity dispersion to be $\sigma^{2} \propto \frac{GM(r)}{r}\propto G n_{0} \mstar r^{2+\alpha}$ and use the same formalism (Eq.~\ref{F17}) to derive Eq.~\ref{F19}.
The capture rate for $r_{\text{crit}}>r_{H}$ becomes:
\begin{equation}\label{F20}
\dot{C} \propto r_{\text{cap}}r_{H}n_{0}^{\frac{3}{2}}G^{\frac{1}{2}}\mstar^{\frac{1}{2}} \left(r_{\text{cap}}r_{H}n_{0} \right)^{-\frac{2+3\alpha}{2\left(1+\alpha \right)}}.
\end{equation}
For simplicity we adopt the number density profile to be $n(r\leq r_{\text{crit}})=n_{0}r^{\alpha}$ and thus assume $\alpha$ to remain constant. Real galactic nuclei with SMBHs more massive than $\unit[10^{7}]{\msun}$ can deviate from
pure power law profiles at radii $r\leq r_{\text{crit}}$, whereas the inner density profiles of large elliptical galaxies are nevertheless well approximated by simple power law profiles \citep{Trujillo2004}. 
Eq.~\ref{F20} is valid for $-3<\alpha<-1$.\\
 
These equations which are derived from the very general angular momentum diffusion concept of \cite{Rees1976}, will lose their applicability for systems where the stellar phase space is not well-occupied with sufficient 
amounts of low angular momentum stars. Gaps in the phase space distribution, for example carved out by binary-SMBH evolution must first be repopulated, whereas the relaxation driven refilling process may take longer than one Hubble time $H_{0}^{-1}$ for large elliptical galaxies.  
Hence for these systems the two-body relaxation driven capture rate will be strongly suppressed \citep{Merritt&Wang}.

\section{Description of the N-body models}\label{NBODY}
In the following sections the computations and results will be specified. 
We make use of conventional N-body units \citep{Heggie1986}. For readers being unexperienced
with these units, a very short overview is given below.  

\subsection{N-body units}\label{3.0}
The set of N-body units is defined by
\begin{equation}
G=M=1
\end{equation}        
where $G$ is the gravitational constant and $M$ the total mass. If the system 
is gravitationally bound and in virial equilibrium with $r_{\text{vir}}=1$ then the total energy $E$, which is the sum of the kinetic and potential energies of all particles, is $E=-\frac{1}{4}$.
N-body timescales which are used as the time base for the computations are defined to be $\frac{t_{\text{cross}}}{2\sqrt{2}}$. Here $t_{\text{cross}}=\frac{2r}{\sigma}$
is the crossing time of the particles at $r=r_{\text{vir}}$. The half mass i.e. half light radius $r_{e}$ for a constant $\frac{M}{L}$-ratio is usually scaled to equal $r_{e}=r_{\text{vir}}=1$. It 
is of the order of kpc scales in physical units for real elliptical galaxies.
For example one N-body timescale would correspond to $t = \unit[7\cdot 10^{6}]{\text{yr}}$ in physical units for a spherical bulge or galaxy of S\'{e}rsic type (n=4, see~\S~\ref{3.1}) with a half light radius $r_{e}\approx \unit[0.65]{\text{kpc}}$
and total stellar mass $M= \unit[10^{9}]{\msun}$. 
In \S~\ref{5.1} and Appendix~\ref{AppendixB} the detailed procedure how the computational results are transformed from N-body to physical units is given.

\subsection{Generation of the models}\label{3.1}
\begin{figure}
\begin{center}
\includegraphics[width=8.5cm]{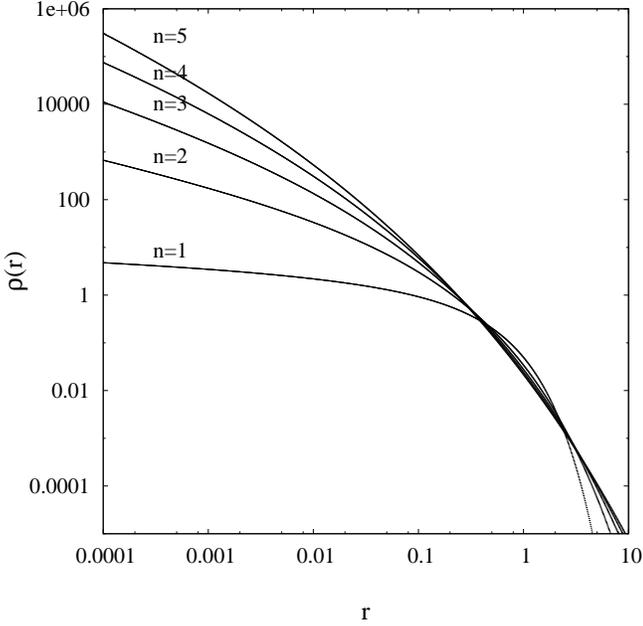}
\end{center}
\captionsetup{format=plain,labelsep=period,font={small}}
\caption{Scale-free density profiles of different S\'{e}rsic models.}
\label{sersicdensity.eps}
\end{figure}
The observed surface brightness profiles $I_{n}(r)$ of bulges and elliptical galaxies are well approximated by the following S\'{e}rsic law \citep{Sersic1968, Caon1993}:
\begin{equation}\label{F49}
I_{n}(r)=I_{e}\exp \left\{-b_{n}\left[\left(\frac{r}{r_{e}}\right)^{\frac{1}{n}}-1\right]  \right\}.            
\end{equation}
Here $n$ is the S\'{e}rsic index. It represents the strength of light concentration towards the center. The parameter $I_{e}=I(r_{e})$ specifies the surface brightness
at the corresponding half light radius $r_{e}$, whereas $b_{n}$ is a scaling factor \citep{Ciotti1999}. The 2D density profile can also be reconstructed from the measured surface brightness profile
using an appropriate mass-to-light ratio, which for our purposes is assumed
to be constant along the radial distance to the center of the galaxy. \\

In order to study the environmental impacts of massive black holes, an unaltered and original S\'{e}rsic density profile is chosen for the initial state of the models.
These N-body models are set up using the same method as described in \cite{Hilker2007}. First the 2D S\'{e}rsic models are deprojected into 3D density distributions using
Abel's integral equation. From the 3D density profile $\rho(r)$, the potential, $\phi(r)$, and mean mass within radius $r$, $M(<r)$, can be deduced.
For a non-rotating, spherical system with an isotropic velocity distribution, the distribution function $f(H)$ is ergodic, where $H$ is the Hamiltonian i.e. the total
energy of the system. The radial velocity distribution is then derived from $f(H)$ by using eq.~4.46a from \cite{Binney2008}. The actual positions as well as velocities of the N-body particles are distributed 
correspondingly in 3D. 
The program is modified by adding a $1/r$-potential of the black hole of mass $\mbh=0.01$ in N-body units \citep{Heggie1986}\footnote{The differences between a $1/r$-potential and a realistic Schwarzschild or Kerr black hole potential
are completely insignificant for distances of several hundred $r_{\text{cap}}$ away from the black hole. This is typically the distance where the innermost particles are located.}.
This step is necessary because otherwise the velocities of particles close to the black hole in the N-body computations would be too low and the system out of equilibrium.
The cut off radius for the models is chosen to be 20 times the half light/mass radius. \\ 

\subsection{NBODY6 numerical dynamics software}
The up to date version of NBODY6 \citep{Aarseth1999, Aarseth2003} with Graphical Processing Unit (GPU) support is used for the direct N-body integrations. 
A black hole is added by a SMBH particle of mass $\mbh=0.01$. It is implemented into the models at the center of mass while being initially at rest.
Particles which fall below the limit of the capture radius $r_{\text{cap}}^{\text{sim}}$ are removed from the simulations while their masses are added to the SMBH particle. The capture radius $r_{\text{cap}}^{\text{sim}}$ remains
constant\footnote{In reality the capture radius would change as well as the total number of capture events. However in order to simplify our extrapolation formalism to realistic galaxies and due
to the fact that the mass gain of the black hole within $T=100$ NBODY timescales is limited to the order of a few percent, it is assumed to be constant.}. To ensure correct dynamics, the SMBH particle receives the center of mass velocity after the capture event.    

\begin{figure*}
\begin{center}
\includegraphics[width=18cm]{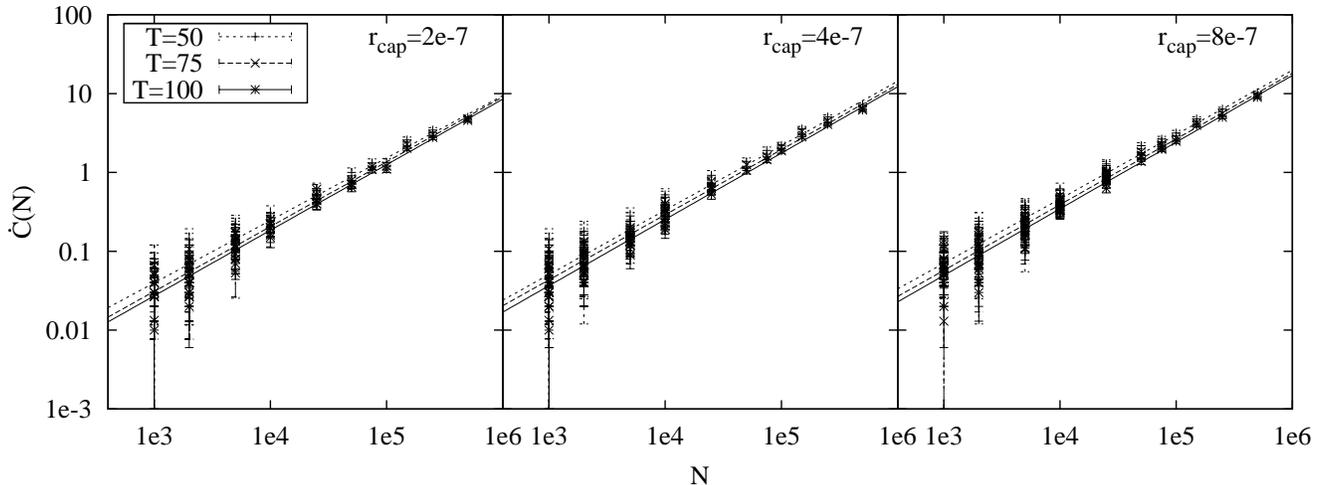}
\end{center}
\captionsetup{format=plain,labelsep=period,font={small}}
\caption{The capture rates per one N-body time unit for the three different black hole capture radii, evaluated from the total amount of swallowed particles within the
 timespan of $T=50,75  \ \& \ 100$ N-body time units.
These values are best fitted by the power law function $\dot{C}(N)=aN^{b}$, here $N$ refers to the total number of simulated particles.}
\label{Capture.ps}
\end{figure*} 

\subsection{The need for a large set of simulations}\label{3.2}
In order to extrapolate many scale-free models to astrophysical systems which contain some orders of magnitudes more stars than are possible to be simulated with
direct N-body integration methods on modern GPUs, the relaxation driven effects in angular momentum and energy space as well as every other $N$ dependent systematic effect (see~\S~\ref{3.3} \& \S~\ref{3.4}) must be
determined. This can be achieved by simulating models with different numbers of 
particles but otherwise identical physical parameters. In doing so several particle models following a 
S\'{e}rsic $n=4$ density profile are generated. It is desirable to simulate these models for as many different black hole configurations as possible in order to use the formalism in \S~\ref{5.1}
for the extrapolation to the black hole of interest, hence increasing the computational effort considerably.
The masses of the particles $m_{i}=N^{-1}$ are always scaled to ensure $\sum_{i}m_{i}=1$ in N-body units \citep{Heggie1986}. N=15\unit[$\times1$]{$\text{k}$}, 15\unit[$\times2$]{$\text{k}$},
10\unit[$\times5$]{$\text{k}$}, 5\unit[$\times10$]{$\text{k}$}, 5\unit[$\times25$]{$\text{k}$}, 2\unit[$\times50$]{$\text{k}$}, 2\unit[$\times75$]{$\text{k}$} and one model containing each \unit[$100$]{$\text{k}$}, 
\unit[$150$]{$\text{k}$}, \unit[$250$]{$\text{k}$} and \unit[$500$]{$\text{k}$} particles are generated and simulated.
All these models are simulated forward in time up to $\frac{100}{2\sqrt{2}}$ crossing times at the virial radius $r_{\text{vir}}=1$ i.e. 100 N-body timescales for three different black hole
capture radii $r_{\text{cap}}^{\text{sim}}=2,4,8\cdot10^{-7}$. 
Energy values and relative energy errors $\left| \Delta E \right| =\left| \frac{E(t_{n})-E(t_{n-1})}{E(t_{n-1})} \right|$ are evaluated directly with the NBODY6 software and controlled every new N-body timescale.
The relative energy errors usually not exceeded values of $|\Delta E|=10^{-8}-10^{-4}$. A few models 
had to be discarded afterwards as they suffered from repetitive energy errors in excess of $|\Delta E|=10^{-2}$. To guarantee unbiased capture rates we also discarded models in which the position of the
black hole was offset by a distance $d\ge 0.1$ from the density center of the particle distribution.
The statistical significance of the numerous low $N$ models is increased by simulating as many models as possible. The required time for the computations of all simulations
exceeds a timespan of seven months on five modern GPUs.

\section{Results}\label{5}
\begin{figure}
\begin{center}
\includegraphics[width=8.5cm]{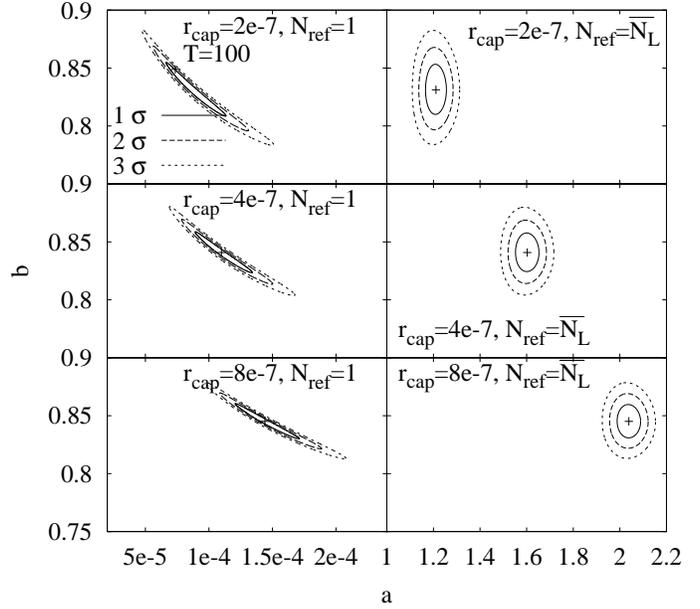}
\end{center}
\captionsetup{format=plain,labelsep=period,font={small}}
\caption{The error ellipses for the models after $T=100$ before (left) and after (right) renormalization. The shape of the error ellipses becomes nearly circular which proofs the parameters to be uncorrelated.}
\label{Errorellipse.eps}
\end{figure}

\begin{figure}
\begin{center}
\includegraphics[width=8.5cm]{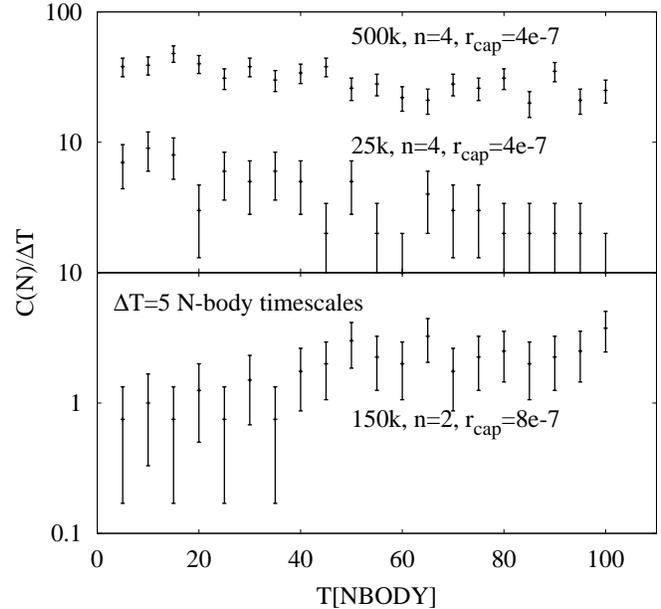}
\end{center}
\captionsetup{format=plain,labelsep=period,font={small}}
\caption{Time evolution of the capture rates for S\'{e}rsic $n=4$ \& $n=2$ models. The statistical significance of the latter ones is increased by averaging over three simulations.}
\label{captime.eps}
\end{figure}

In Fig.~\ref{Capture.ps} the number of particles being swallowed by the black hole is plotted against the total number of particles. This is done for each black hole
capture radius $r_{\text{cap}}^{\text{sim}}=2,4,8\cdot10^{-7}$. Moreover the total number of captured particles within $T=50, 75, 100$ N-body integration times is divided by these values to obtain the capture rate per N-body timescale. 
The number of captures averaged over all runs are then
approximated by a power law function 
\begin{equation}\label{FP}
\dot{C}(N)=aN^{b} 
\end{equation}
with the help of the \textit{Marquardt-Levenberg} minimization method and independently by a grid scanning algorithm minimizing the Chi-square error statistics.
The free parameters $a$ and $b$ have to be positive real numbers while the boundary condition $\dot{C}(N)|_{N=0}=0$ requires no offset.
To reduce the correlation between the parameters $a$ and $b$ to zero, we normalize the power law function $\dot{C}(N)=a^{\prime}\left(N / \bar{N}_{L}\right)^{b}$ during fitting. The denominator $\bar{N}_{L}$ refers to
the logarithmic mean. The resulting effect can be seen in Fig.~\ref{Errorellipse.eps}. These uncorrelated values\footnote{To simplify the extrapolation formalism, the 
renormalized constant of proportionality $a^{\prime}$ and its error is afterwards transformed back to $a=\frac{a^{\prime}}{\bar{N}_{L}^{b}}$. This does not affect the correlation coefficient $\rho=0 $ between $a,b$.} are used 
for the extrapolation to realistic values.
The justification for using a power law approximation for the 
capture rate $\dot{C}(N)$ from the simulations comes from Eq.~\ref{F18} when replacing $n_{0}=N\rho_{0}$ and $\mstar=N^{-1}$. 
Poisson square root errors $\sqrt{N_{c}}$ are assumed for all values and ${N_{c}}$ is the total number of captured particles. The results can be found in Table~\ref{tablen=4}.
Additionally the reduced Chi-Square values $\chi_{\mu}$ and the $\chi^{2}$-probability function $Q(\mu,\chi^{2})$ are calculated 
in order to test the validity of a power law approximation for the capture rate. Given the values in Table~\ref{tablen=4}, the hypothesis of a power law function seems to be
a reasonable assumption. However, for the determination of the error values of parameters $a,b$ the square root errors $\sqrt{N_{c}}$ are rescaled slightly by the values $\sqrt{\chi_{\mu}}$ from Table~\ref{tablen=4}
to obtain $\chi_{\mu}=1$. Otherwise the quoted error values would be underestimated for the case of $\chi_{\mu}\ge1$ \citep{Press1992}\footnote{Chapter 15.1}.  \\ 
\begin{table}
%\begin{center}
  \begin{tabular}{|c|c|c|c|c|c|}
    \hline
    T & $r_{\text{cap}}^{\text{sim}}$ & $a(10^{-5})$ & $b$ & $\chi_{\mu}$ & $Q$\\ 
    \hline
  50  & $2\cdot 10^{-7}$   &$16.76 \pm 3.11$ & $0.792\pm0.018$ &0.82   & 0.792\\
    &$4\cdot 10^{-7}$   &$ 17.95  \pm 2.89 $ & $ 0.817\pm 0.014 $& 1.06 & 0.359 \\
    &$8\cdot 10^{-7}$   &$25.08\pm3.92$ &$0.816\pm0.014 $&1.41 & 0.024\\
    \hline

   75 &$2\cdot 10^{-7}$   &$ 10.63 \pm 1.95$ &$0.822 \pm 0.016$& 1.01 & 0.450\\
    &$4\cdot 10^{-7}$   &$  14.46  \pm  2.09 $  &$0.826  \pm 0.013$& 1.13& 0.245 \\
    &$8\cdot 10^{-7}$   &$18.15 \pm 2.51$ &$0.833\pm0.012$& 1.41 & 0.022\\
    \hline

  100  &$2\cdot 10^{-7}$   &$ 8.73 \pm 1.33$ &$0.831\pm 0.013$& 0.83 & 0.788 \\
    &$4\cdot 10^{-7}$   &$10.98 \pm  1.49$   &$0.841 \pm 0.012$& 1.12 & 0.255\\
    &$8\cdot 10^{-7}$   &$14.41 \pm 1.97$ & $0.845\pm 0.012$&1.57 & 0.005\\
    \hline
\end{tabular}
\captionsetup{format=plain,labelsep=period,font={small}}
  \caption{Fit parameters of the power law approximation (Eq.~\ref{FP}) for the simulated S\'{e}rsic $n=4$ models. The black hole capture radii and timescales $T$ are given in N-body units.
$\chi_{\mu}=\chi^{2}/\mu$ corresponds to the reduced Chi-Square values, $\mu$ are the degrees of freedom and $Q=\Gamma(0.5\mu,0.5\chi^{2})$ the $\chi^{2}$-probability function which estimates the 
likelihood of the power law fit.}
\label{tablen=4}
%\end{center}
\end{table}
The advantage of numerical simulations over analytical expressions like Eq.~\ref{F18} are given in the ability to take dynamical aspects like cusp formation, dynamical heating (\S~\ref{3.3}) and a wandering SMBHs (\S~\ref{3.4}) into account.
These depend strongly on time and on the total number of particles and may influence the capture rate $\dot{C}(N)$. The predicted power-law index of Eq.~\ref{F18} is therefore not expected to exactly match the value obtained from the simulations. 
In fact Eq.~\ref{F18} would only predict $\dot{C}(N)\propto N^{\frac{4.5+2\alpha}{4+\alpha}}\approx N^{0.6}$ for $\alpha \approx -1.5$ compared to $\dot{C}(N)\propto N^{0.83}$ from the  
computations. The difference is caused by stronger dynamical evolution and cluster heating in low number particle simulations accompanied by a decrease in the total number of particles falling into the black hole. Models
containing many more particles have much smoother potentials and relaxation driven effects (notably cluster heating) need longer to influence (decrease) the capture rate (Fig.~\ref{captime.eps}).
Consequently the exponent $b$ of the power law function which approximates the number of captured particles of the total set of simulations becomes
larger than expected from Eq.~\ref{F18}. These dynamical processes are reflected by the values of $a,b$ at different timescales. 
The constant of proportionality $a$ decreases in time, whereas the slope parameter $b$ is consistent with a small increase from $b\approx0.80$ at time $T=50$ up to $b\approx0.83$ at time $T=75$.  
Thus the exponent of the power law function which approximates the capture rates becomes
slightly larger, whereas the constant of proportionality decreases. Moreover the $T=50$ values may still be influenced by initial conditions. There are minor changes in $b$ from $T=75$ to $T=100$. \\

For the purpose of this study the rate $\dot{C}$ is assumed, within the statistical uncertainty, to remain unchanged when extrapolated to larger values of N. This assumption can only hold if the phase space is already well   
occupied with sufficient amounts of low angular momentum stars. This is a necessary condition for the steady diffusion process of stars into the loss cone.
The capture rates of the S\'{e}rsic $n=4$ models are found to be maximal at the beginning of
the simulations in contrast to S\'{e}rsic $n=2$ models with their much shallower density profiles (Fig.~\ref{captime.eps}).
This demonstrates the above assumption to be credible, at least for galactic nuclei containing SMBHs less massive than $\unit[10^{7}]{\msun}$. In such galaxies the diffusively refill of any small gap with radius 
$r_{\text{gap}}<<r_{H}$ would anyway occur on a timescale shorter than a Hubble time \citep{MerrittDM2005}.
The observed strong $N$ dependence (b=0.83) may become irrelevant or absent for black holes more massive than $10^{7}\msun$, especially if they have core profiles. For these systems 
the loss cone refilling timescale $T_{\text{refill}}\approx \theta_{\text{lc}}^{2}T_{\text{rel}}$, becomes very long. Once the initially filled loss cone becomes emptied within a few crossing times,
the capture rate $\dot{C}$ would stagnate at insignificant values as long as there is no re-population mechanism more efficient than angular momentum diffusion \citep{Merritt&Wang, MerrittDM2005}.\\

This effect can be illustrated by simulating S\'{e}rsic $n=2$ models. These have a slower dynamical evolution, a different cusp and cluster heating timescale and a reduced population of low angular momentum
stars compared to the S\'{e}rsic $n=4$ models. In this way, qualitative limitations on the number of capture events for core-type galaxies with shallow central density profiles (Fig.~\ref{sersicdensity.eps})
can be obtained. Even though the extended outer profiles of the 
most-massive elliptical galaxies are conform with a large S\'{e}rsic index $n$, the 'depleted' core-type central regions (this is where the relevant black hole physics take place) are more similar in their appearance to the shallow centers
of low $n$ models\footnote{This is also one reason which complicates the discrimination between 'true' cores, formed by the dynamical evolution of massive binary black holes 
and those in which only the outer envelopes are modified by near encounters. In the latter case the outer profile
extrapolated to inwards radii would suggest the existence of a core \citep{Hopkins2010}. The binary black hole 
mechanism may also be accompanied by other processes lowering the central stellar density \citep{Merritt2010, Schawinski2006}.}.   
A strongly reduced disruption rate in comparison 
to the S\'{e}rsic $n=4$ models is evident in these computations. The enlarged radius of influence $r_{H}$ and therefore the difference in the extrapolation formalism to realistic galaxies can not 
compensate these differences. Moreover in the largest simulated S\'{e}rsic $n=2$ models, the capture rate stagnate first around insignificant values. It starts increasing (Fig.~\ref{captime.eps}) afterwards,
accompanied by the relaxation driven formation of a cusp and a population of stars with sufficiently low angular momentum. 
If we assume this behaviour to persist unchanged up to even larger numbers of particles, i.e. to large core-type galaxies where no cusps can form on timescales shorter than $H_{0}^{-1}$, these numerical findings confirm analytical predictions \citep{Wang2003} in a qualitative way.
The capture rate of stars 
in large core-type galaxies is very low, as long as the diffusive refill of the angular momentum space with a sufficient number of stars, i.e. the cusp formation timescale, takes longer than a Hubble time.\\
See also \S~\ref{5.3} \& Appendix~\ref{AppendixC} for more details on this topic. \\
 
Finally the here performed simulations of the S\'{e}rsic $n=4$ models strongly support the scenario of \cite{Rees1976} in which stars are driven into SMBHs via diffusion in 
angular momentum space and not only by diffusion in energy space. From the most elementary considerations of energy diffusion and by assuming the two-body relaxation time to be
$T_{\text{rel}}\propto \frac{N}{ln(N)}$, one would expect $\dot{C}(N)=\frac{\text{d}N}{\text{d}t}\propto N\cdot T_{\text{rel}}^{-1}\propto \ln(N)$. 
Such a small increase of the capture rate with N is incompatible with our results. 

\section{Dynamics \& Scaling issues}\label{3}
The capture rate is influenced by several dynamical processes which are described below.         
\subsection{Cusp formation and cluster expansion}\label{3.3}
The process of relaxation strongly influences the dynamics of stars around a SMBH. In \cite{Bahcall1976} the relaxation driven
evolution of the stellar density profile near a SMBH is determined.
It is found that the energy which some stars loose through near encounters is balanced by an outgoing flux of energy if the slope of the density profile
is $\alpha=-1.75$. The required time to form such an equilibrium density B\&W profile strongly depends on the relaxation time which becomes larger the smoother a gravitational potential is \citep{Spitzer1987}. \\
\begin{figure}
\begin{center}
\includegraphics[width=8.5cm]{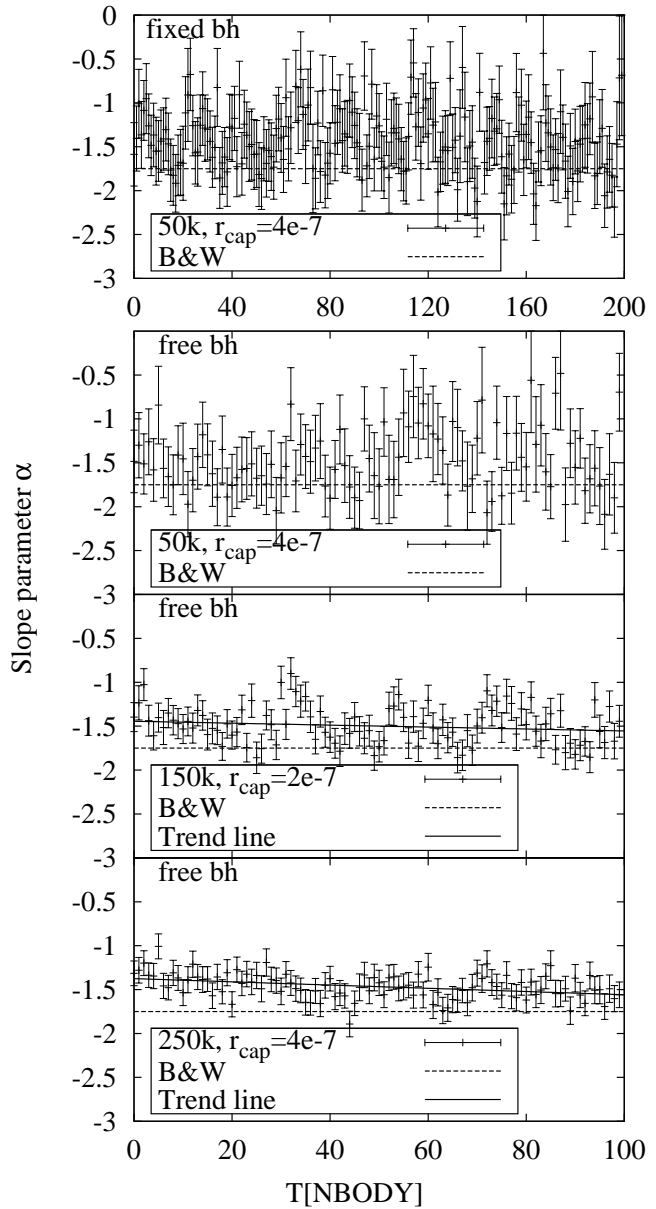}
\end{center}
\captionsetup{format=plain,labelsep=period,font={small}}
\caption{Time evolution of the central slope parameter $\alpha$ plotted for the \unit[$50$]{$\text{k}$}, \unit[$150$]{$\text{k}$} and \unit[$250$]{$\text{k}$} models. The linear trend (solid line) is only drawn for the \unit[$150$]{$\text{k}$} and \unit[$250$]{$\text{k}$} models, while
the first one (fixed black hole) was simulated forward in time up to $T=200$.}
\label{slope.eps}
\end{figure} 

\begin{figure*}
\begin{center}
\includegraphics[width=18cm]{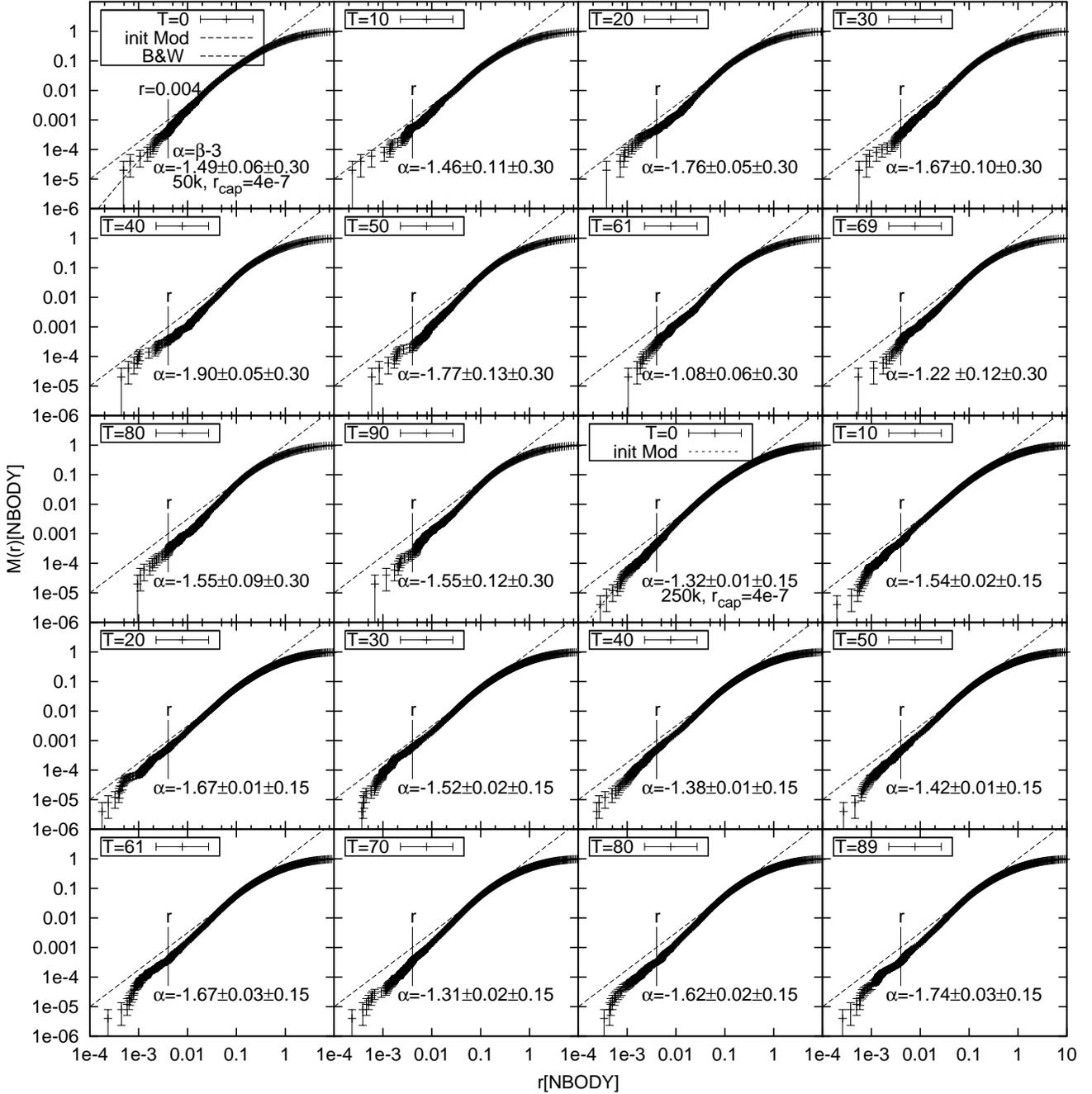}
\end{center}
\captionsetup{format=plain,labelsep=period,font={small}}
\caption{Mass profiles of two models. The thin dashed black line represents the gradient of the B\&W profile while the thick dashed black line (only drawn for T=0) displays the unaltered S\'{e}rsic $n=4$ model. The first 
error on $\alpha$ corresponds to the fitting error while the second one to the statistical error inferred from Monte Carlo simulations. The profiles and thus $\alpha$ are evaluated for radii $r\le0.004$. For more informations see the text below.}
\label{massdensity.eps}
\end{figure*} 

The $\alpha=-1.75$ profile is compatible with the present N-body models only up to $N=\unit[25-50]{\text{k}}$ where relaxation is strongest and
the statistical scatter is large. The $N>\unit[50]{\text{k}}$ models, which allow a more precise measurement of $\alpha$, are found to be in the developing stage towards more cuspy profiles.
In Fig.~\ref{slope.eps} the time dependent central slope parameter $\alpha$ within $r=0.004$ is plotted for some models. The radius $r$ is chosen to be $20\%$ smaller than the time and model-averaged black hole influence
radius\footnote{See Appendix~\ref{AppendixB} for information regarding the determination of $r_{H}$.} in order to ensure that the slope parameter is not determined for radii larger than $r_{H}$ at the beginning of the simulation
when the mass and influence radius of the black hole is smallest.  
In order to obtain the central slope parameter $\alpha$, it is inappropriate to calculate the density profile $\rho(r)\propto r^{\alpha}$ from given shells of thickness $\Delta r$ and densities $\rho(r+\Delta r)$.
Unfilled shells, especially in low N models, would strongly bias the 
determination. In order to circumvent this difficulty, the cumulative mass function $M(r)\propto r^{\beta}\propto \int_{0}^{r^{\prime}}r^{2}\rho_{0}r^{\alpha}\text{d}r$ is calculated and the density slope parameter 
$\alpha=\beta-3$ (equating coefficients) is determined from the measured $\beta$. This approach is tested by Monte-Carlo simulations in which several thousand models of particles following a $\rho\propto r^{-1.75}$ distribution are realized.
For each of these models 
the central slope parameter $\alpha$ within $r=0.004$ is calculated. The models are scaled such that the number of particles within $r=0.004$ is equal (within the statistical scatter) to those of the 
\unit[$25$]{$\text{k}$}, \unit[$50$]{$\text{k}$}, \unit[$75$]{$\text{k}$}, \unit[$150$]{$\text{k}$} and \unit[$250$]{$\text{k}$} simulations. The 
standard deviation $\sigma$ from the obtained normal distribution\footnote{Actually very small particle numbers within $r=0.004$ bias the power-law density-approximation and the distribution of central slope parameters becomes
asymmetric with a tail towards very large values. This may partially account for some extreme outliers especially in low N models, whereas for larger models the distribution becomes more symmetric and the expectation values $\mu$ center
around $\alpha=-1.75$.} of central slope parameters is then taken as a reasonable estimate for the statistical error in addition
to the one obtained from the fit itself. In Fig.~\ref{massdensity.eps} the time evolution of the mass profiles of two models are plotted.  \\ 
\begin{figure}
\begin{center}
\includegraphics[width=9.5cm]{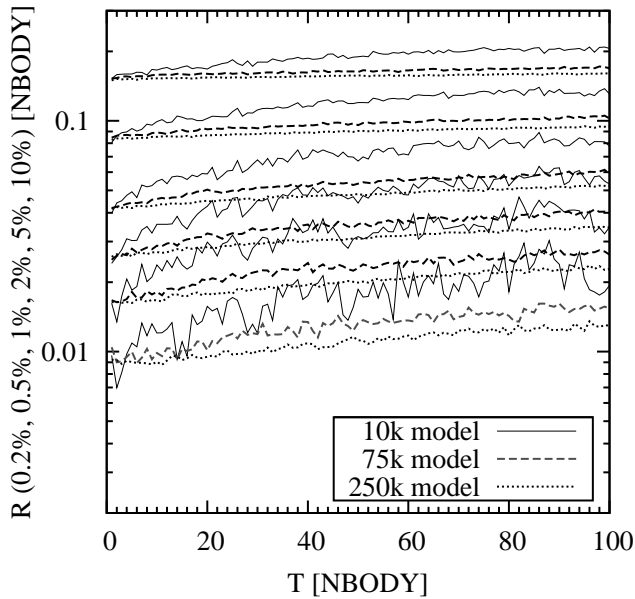}
\end{center}
\captionsetup{format=plain,labelsep=period,font={small}}
\caption{A comparison between the time evolution of several Lagrange radii for three different simulated models. As expected from theory, the Lagrange radii  
evolve faster to larger values in simulations containing fewer particles. The fluctuations are statistical in nature. The position of the black hole is used as the reference center.}
\label{Lagrangeradii.eps}
\end{figure}

In order to estimate the dependence of a wandering black hole on cusp formation processes and finally the capture rate, simulations of fixed black holes are desirable.
Such simulations are realized by making use of a modified NBODY1 code (see \S\ref{3.5} for more details regarding the capture rates).
Within the large statistical errors, no significant difference in the density profiles between the free floating and fixed black hole is identified for the \unit[$50$]{$\text{k}$} model. 
This is not an unexpected finding since the most bound particles, which are also the particles with the highest probability of being captured, are
expected to follow the motion of the black hole. However a rigorous statistical evaluation is beyond the scope of this study.\\

While the capture rate is increased by cusp formation, dynamical heating counteracts by reducing the central density. The cluster starts to expand by decreasing the absolute value of its binding energy due to 
increasingly more strongly bound particles
which are losing energy by relaxation. These particles, which may finally be swallowed by the black hole, are transferring their kinetic energy to other particles.  
This heating is illustrated by the time evolution of the Lagrange radii (Fig.~\ref{Lagrangeradii.eps}).
As a consequence the capture rate is expected to depend strongly on the density profile close to the black hole (Eq.~\ref{F18}).\\

In reality mass segregation of heavier bodies being relevant for multi-mass systems \citep{Alexander2009b, Baumgardt2004b, Morris1993, Preto2010}, stellar collisions \citep{Bailey1999, Dale2009}, a significant fraction of primordial
binary stars \citep{Hopman2009}, torques from anisotropic matter distributions acting as massive perturbers \citep{Perets2007}, star formation
by gas inflow \citep{Hopkins2010b} and the possible presence of IMBHs \citep{Baumgardt2006b} would complicate the dynamics of stars close
to a SMBH even more. These effects are also expected to accelerate the dynamical evolution and to enhance the number of stellar disruption events. Newly formed
stars may replace those lost by tidal disruptions while tidal torques from IMBHs or a second SMBH are expected to
refill the loss cone efficiently. Recoiled black holes should also enforce a burst of stellar disruptions \citep{Stone2010}. 
In nature the relaxation driven B\&W cusp formation takes very long and is expected to exceed one Hubble time $H_{0}^{-1}$ for black
hole masses larger than $\unit[10^{7}]{\msun}$ \citep{Freitag2008}.\\

\subsection{Wandering black hole}\label{3.4}
In the simulations the SMBH particle responds to the interaction with other particles which causes the SMBH to wander. This might affect the formation of a density cusp and influence the capture rate \citep{Baumgardt2004a}. 
\cite{Chatterjee2002} gives  a very detailed overview of the relevant forces acting on a SMBH. They are summarized below.\\

\begin{figure}
\begin{center}
\includegraphics[width=8cm]{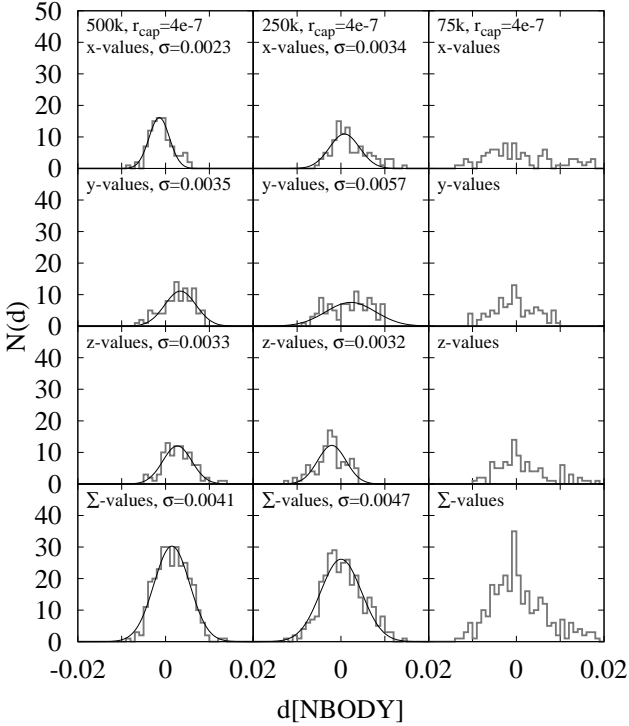}
\end{center}
\captionsetup{format=plain,labelsep=period,font={small}}
\caption{The 100 binned ($\Delta d=0.001$) x, y, z-positions of the SMBH particle with a capture radius $r_{\text{cap}}^{\text{sim}}=4\cdot10^{-7}$ for the \unit[$500$]{$\text{k}$}, \unit[$250$]{$\text{k}$} and \unit[$75$]{$\text{k}$} model.
 In the last row the sum of these values is plotted and approximated by a normal distribution.  
 The probability distributions are well approximated by a Gaussian underlining the character of the
Brownian motion. The relevant length scale $d$ is given in units of the virial radius $r_{\text{vir}}=r_{e}=1$. The SMBH particle in the \unit[$75$]{$\text{k}$} model experienced a minor kick during the integrations.}
\label{Maxwell}
\end{figure}

The here performed simulations differ only in two ways from the N-body simulations done
by \cite{Chatterjee2002}. The black hole is allowed to swallow particles and the forces are unsoftened. 
The SMBH moves around the common center of mass due to the gravitational interaction with particles bound to it, whereas unbound particles are forcing
the black hole to wander in a way which resembles the Brownian motion of molecules. The latter process is the dominant contribution to the wandering of the black hole (see Fig.~\ref{Maxwell}).\\

The situation is now complicated by the possible occurrence of violent three body encounters, e.g the interaction between the black hole, a strongly bound particle in orbit around it 
and another one. Recoil events force the black hole and its
surrounding particles to move outwards. 
The mass fraction, $\frac{m}{\mbh}$, is usually orders of magnitudes larger in any performed N-body simulation than it is in a 
realistic nucleus of a galaxy. And, because the recoil effect becomes stronger for a larger fraction $\frac{m}{\mbh}$ and for smaller capture radii $r_{\text{cap}}^{\text{sim}}$, wandering of the SMBH in the N-body models
is expected to modify the processes leading to the formation of a cusp. Consequently, the simulated capture rate may become affected.
If the recoil kick of the SMBH particle is strong enough to eject it out of the density center or even from the whole cluster, the capture rate would drop
significantly. This is expected, due to obvious reasons, to happen more likely in simulations with low particle numbers. 
As a consequence the extrapolated N-dependent capture rate would be strongly biased and the best fitted slope parameter, b, may be too large.
Therefore the actual position of the SMBH particle is compared to the density center of the matter distribution for every simulated model and at every new N-body time unit.
The black hole particle is not considered in the calculation of the density center which is determined by the method described in \cite{Casertano1985}.
If the position of the black hole and the density center are offset from each other by $d=0.1$ in N-body units, the simulation is removed and replaced by a different one. 
In nearly all simulations this offset is smaller than $10^{-3}-10^{-2}$. This guarantees that the results are not biased by displaced black holes in the low N models.\\

\begin{figure}
\begin{center}
\includegraphics[width=8cm]{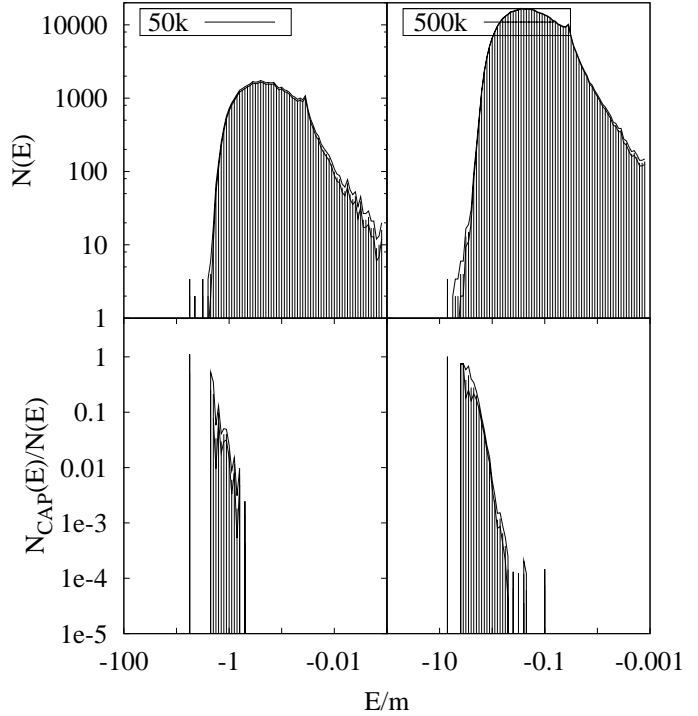}
\end{center}
\captionsetup{format=plain,labelsep=period,font={small}}
\caption{In the upper two figures the binned specific energy $\frac{E}{m}$ distributions of the \unit[$50$]{$\text{k}$}, $r_{\text{cap}}^{\text{sim}}=4\cdot10^{-7}$ and \unit[$500$]{$\text{k}$}, $r_{\text{cap}}^{\text{sim}}=4\cdot10^{-7}$
models are plotted. The lower diagrams depict the ratio of captured particles to total number of particles within the given energy bins. Evidently only the particles with the most negative energy i.e., the most strongly
bound particles are accreted as expected from theory. The upper and lower (black) lines represent the error uncertainties.}  
\label{energy.eps}
\end{figure} 

But even by removing those few models where "unnatural`` kicks and displaced black holes are observed, the wandering of the black hole itself might affect the capture rate. The wandering radius can be 
determined by the standard deviation of the normal distribution (Fig.~\ref{Maxwell}). It is found to be comparable in size 
to the influence radius $r_{H}=0.005$ (for the \unit[$250$]{$\text{k}$} model) and becomes gradually smaller for larger particle numbers i.e. smaller mass fractions $\frac{m}{\mbh}$.  \\
 
A first clue about the degree to which the wandering black hole affects the results can be obtained by a closer look at the energies of accreted particles. If only particles are swallowed which are strongly bound i.e. have the most negative
energies, the effect of Brownian motion on the capture rate is expected to be rather small, since the cloud of strongly bound particles moves together with the black hole. In Fig.~\ref{energy.eps} the initial energy distribution
for two models is shown. Also plotted is the fraction of the accreted particles to the total number of particles within a given energy bin. Evidently only the most strongly bound particles are captured. 
If the energy $E=-\frac{\mbh m}{r} + 0.5mv_{m}^{2}+0.5\mbh v_{\mbh}^{2}$ of the particle of mass $m$ and black hole is negative, shortly before it enters the capture radius and is removed, the particle is 
gravitationally bound to the SMBH. In our models the vast majority of particles are gravitationally bound to the black hole, e.g. the fraction of bound particles centers around 100\% in the low-N models  
and 85 - 95\% in the largest-N models.   \\

We therefore conclude that a wandering black hole does not bias the capture rate in a way that would make it unrealistic when extrapolated to real IMBHs and SMBHs. The performed simulations automatically contain
the gradual change in the number of accreted particles which are influenced by the wandering of the black hole. Our largest N-computations
already approach realistic IMBHs embedded in globular clusters. To resolve all doubts that the steep dependence on N of the capture rate, $\dot{C}\propto N^{0.83}$, is not caused by the systematics of the wandering black hole, especially
in low N models, direct N-body simulations with fixed black holes (\S~\ref{3.5}) are performed.  
For completeness it should also be mentioned that the N-body models include two additional effects: (i) A restoring force which arises between the black hole and the overall potential of the stellar distribution, especially
if it has a cuspy density center, and (ii) a dynamical frictional force when the black hole passes through the cloud of particles \citep{Chandrasekhara, Chandrasekharb,
Chandrasekharc, Chatterjee2002}.\\

\subsection{Fixed black hole}\label{3.5} 
\begin{figure}
\begin{center}
\includegraphics[width=8cm]{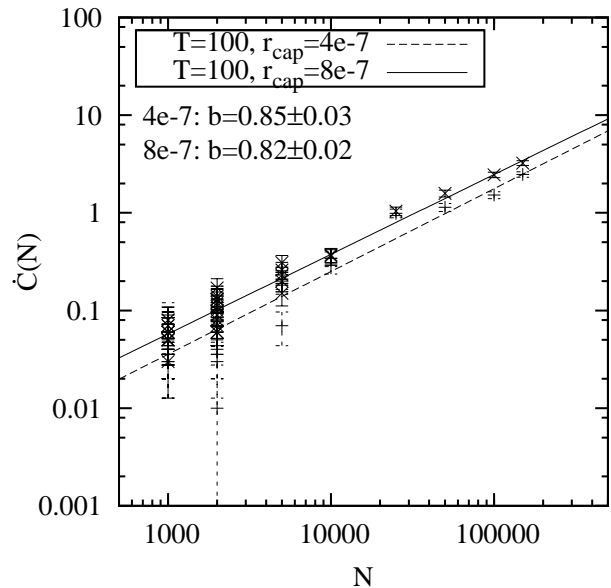}
\end{center}
\captionsetup{format=plain,labelsep=period,font={small}}
\caption{Results of the simulations with a fixed black hole. Strongly bound particles which needed to be removed artificially to prevent slow downs or a computational crash are not considered
in the evaluation of the slope. Note the excellent agreement with the results obtained for the more realistic NBODY6 computations (Fig.~\ref{Capture.ps}).}  
\label{captureG.eps}
\end{figure}

Simulations with a fixed black holes are realized by using NBODY1. Unfortunately it is impossible in NBODY6 to fix the SMBH particle to a specific location while simultaneously
using all of its computational benefits. On the other hand the usage of an independent N-body software implementation reduces the possibility of systematic errors. The NBODY1 simulations are performed on
special-purpose, GRAPE-6A boards \citep{Fukushige2005} at the stellar Populations and Dynamics Research Group in Bonn. 
The black hole is mimicked by an (unsoftened) external $\frac{1}{r}$ potential which is directly implemented into the code. Particles which cross the capture radius are removed, while their masses are added to the mass
of the black hole. To circumvent collisions between field particles, a small softening parameter $\epsilon=10^{-4}$ is used.
Additionally some strongly bound particles around the external potential are erased artificially (the number corresponds to roughly $30\%$ of the total number of ''true'' capture events) in order to prevent gradual slow downs, large energy errors and/or the complete crash of the simulations. 
The energies of all removed particles are handled carefully to ensure a correct energy output. 
Due to these limitations and the much smaller sample of simulated models, the NBODY1 computations are not used for the extrapolation to realistic galaxies but only for a rough comparison to the much more advanced NBODY6 simulations.
In Fig.~\ref{captureG.eps} the results are plotted.
Despite the large simplifications of the NBODY1 computations, the power law index $b$ of the capture rate, $\dot{C}\propto N^{b}$, agrees, within the statistical uncertainties, closely with the index obtained with the much more 
sophisticated NBODY6 simulations with free moving SMBHs. As a consequence a (strongly) wandering black hole particle does not bias the low N results in a way which would be dangerous when extrapolating these 
to astrophysical systems harboring many more stars than particles in our simulations. Of course this behaviour may change for initial black hole masses different from the one $\mbh(t=0)=0.01$ used in these computations.
\section{Discussion}\label{6}
\subsection{Scaling to realistic galaxies}\label{5.1}
The so-far presented results must be scaled to astrophysical systems in order to infer the rates at which stars are disrupted by central, supermassive black
holes. From the following relation 
\begin{equation}\label{F57}   
\frac{r_{\text{cap}}}{r_{H}}\Big|_{\text{sim}}=\frac{r_{\text{cap}}}{r_{H}}\Big|_{\text{astro}},
 \end{equation}
which must be necessarily fulfilled, the capture radii, $r_{\text{cap}}$, for the corresponding black holes of interest must be determined. In order to scale to astrophysical systems,
we use the $\mbh-\sigma$ relation
from \cite{Schulze2009},
\begin{equation}\label{F57a}
\left(\frac{\mbh}{M_{8}}\right)=1.51 \left(\frac{\sigma}{200 \text{km s$^{-1}$}}\right)^{4.32},
 \end{equation}
and the expression for the radius of influence,
\begin{equation}\label{F57b}
r_{H}=\frac{G\mbh}{\sigma^{2}},
\end{equation}
to calculate $r_{H}$ for a SMBH of given mass,
\begin{align}\label{F55}
r_{H} & \approx 13.1\left(\frac{\mbh}{M_{8}}\right)^{0.54}[\text{pc}].
\end{align} 
Here $M_{8}$ corresponds to $10^8\msun$ and for reasons of computational feasibility we neglected the intrinsic scatter of the $\mbh-\sigma$ relation. 
This is useful when dealing with averaged quantities like the impact of stellar disruptions for the growth history of the majority of SMBHs. Some
studies may instead be interested in individual systems and the extrapolation formalism can easily be replaced by direct measurements of $\mbh, r_{H}$ and $\sigma$ 
instead of using the values from the $\mbh-\sigma$ relation. This also holds for   
the choice of the relevant tidal disruption radius,
\begin{equation}\label{F57b1}
 r_{\text{cap}}=g \rstar \left(\frac{\mbh}{\mstar}\right)^{\frac{1}{3}},
\end{equation} 
where $g$ is a parameter depending on the stellar polytrope, mass and spin of the black hole
as well as the trajectory of the star. Black holes below $10^7\msun$ and solar like stars are well approximated by $g\approx 1$ \citep{Kochanek1992}. A more detailed discussion of $r_{\text{cap}}$ can be found in Appendix~\ref{AppendixA}.\\

The relevant astrophysical timescale is obtained through the computation of the crossing time $t_{\text{cr}}(r_{H})=\frac{2r_{H}}{\sigma(r_{H})}$ at the influence radius $r_{H}$ of the black hole in comparison with that of our numerical integrations. The number of
disruption events within the given timescale is obtained from the derived capture rate \footnote{The capture rates $\dot{C}(N)$ from the numerical computations are normalized to one N-body time unit i.e $\frac{1}{2\sqrt{2}}$ crossing time at the virial radius
$r_{\text{vir}}=1$ and must be scaled down to one crossing time at the influence radius of the black hole in order to become synchronized with the astrophysical timescale $t_{\text{cross}}$.}
$\dot{C}(N,r_{\text{cap}}^{\text{sim}})=a(r_{\text{cap}}^{\text{sim}})N^{b}$. Here $N$ refers to the total number of (real) stars with the averaged mass $\mstar$ in the bulge component or whole elliptical galaxy.
It is assumed to be $N=\frac{100\mbh}{\mstar}$ in accordance with our numerical integrations, whereas $a(r_{\text{cap}}^{\text{sim}})$ is extrapolated to the black hole mass of interest by using the $T=100$ values for the parameters $a$\footnote{At least three different
capture radii must be simulated to allow for non linear extrapolation of the parameter $a(r_{\text{cap}}^{\text{sim}})$. This is required for the extrapolation to different black hole sizes/masses.}.
The parameter $b$ is assumed to be unrelated to the capture radius and is hence taken to be constant at $b=0.83\pm0.01$. In fact Eq.~\ref{F18} predicts the slope parameter $b$ to be unrelated to the capture radius. 
Nevertheless a minor change in $b$ towards smaller vales of $r_{\text{cap}}^{\text{sim}}$ cannot be rejected given the $b$ values of Table~\ref{tablen=4} at $T=100$. This might be explained by a combination of timing issues,
simplified assumptions of our analytical approach
or is purely statistical in nature. Therefore the parameter
$b$ is extrapolated (by linear and power law regressions) down to the required values of $r_{\text{cap}}^{\text{sim}}$ in order to test its impact on the capture rates. The impact is found to be moderate 
because $r_{\text{cap}}^{\text{sim}}$ has to be extrapolated down to $r_{\text{cap}}^{\text{sim}}=(0.06-0.07)\cdot10^{-7}$ (depending on the used $\mbh-\sigma$-relation)
for the largest black hole with $10^7\msun$. While the capture rate would be unaffected for the least-massive black holes, it would drop by a factor of $~2$ for the most-massive ones.
Increasing uncertainties of these values due to the propagation of error analysis strongly overlaps with those of fixed $b$. For the purposes of this study we therefore assume the parameter $b$ to be independent of $r_{\text{cap}}^{\text{sim}}$ and
refer the reader to \S~\ref{5.3} for a more critical discussion on that topic as well as of the improvements left for future work.\\

Finally for an individual galactic nucleus hosting a SMBH of mass $\mbh$, with a radius of influence $r_{H}$, velocity dispersion $\sigma(r=r_{H})$, capture radius $r_{\text{cap}}$ and a stellar population with the mean mass
$\mstar$, a very general expression for the capture rate inferred from
the numerical integration can be obtained by applying Eq.~\ref{F57b}:
\begin{equation}\label{F57d}
\dot{C}_{\text{astro}} =0.00061  \left(\frac{\mbh}{\mstar}\right)^{0.951}r_{H}^{-1.363} \rstar^{0.363} g^{0.363} \sigma 
\end{equation}
The validity of Eq.~\ref{F57d} covers the parameter range of IMBHs as well as SMBHs up to $\mbh \approx \unit[10^{7}]{\msun}$. 
In the following section we explicitly make use of the $\mbh-\sigma$ relation and assume only solar like stars as well as $g=1$.

\subsection{Disruption rates of IMBHs \& SMBHs}\label{5.2}
By applying the extrapolation formalism from section \S\ref{5.1}, the integrations yield the following expression for the capture rate of real astrophysical galaxies: 
\begin{equation}\label{F57e} 
 \dot{C}(\mbh)=6.29\cdot10^{-8}\left(\frac{\mbh}{\msun}\right)^{0.446} \left[\text{yr}^{-1}\right].
\end{equation}
For comparison the results are also extrapolated according to an older version of the $\mbh-\sigma$ relation from \cite{Ford2005} to illustrate the dependence of the capture rate on systematic black hole mass determinations:
\begin{equation}\label{F57f} 
 \dot{C}(\mbh)=3.54\cdot10^{-7}\left(\frac{\mbh}{\msun}\right)^{0.353} \left[\text{yr}^{-1}\right].
\end{equation}
These results holds for nonrotating isotropic galaxies or globular clusters with cuspy inner density profiles $\rho(r)\propto r^{\alpha}$, where the density power law index is $\alpha \approx -1.5$. Eq.~\ref{F57e} and~\ref{F57f}
should not be applied to black holes with masses larger than $10^7\msun$. The uncertainties correspond to about $50\%$ of the values (see Table~\ref{tablen=5}).
The interested reader is referred to Appendix~\ref{AppendixB} for a much more detailed description of how the numerical results are extrapolated to realistic galaxies.
The astrophysical disruption rates of stars (including the statistical uncertainties) for some exemplary black holes are summarized in Table~\ref{tablen=5}.
We also calculate the disruption rates for IMBHs in order to compare them with previous
simulations \citep{Baumgardt2004a}.
\begin{table}
\begin{center}
  \begin{tabular}{|c|} 
  \cite{Schulze2009}
   \end{tabular}
    \begin{tabular}{|c|c|c|c|}  
    $ \mbh (10^{6}{\msun})$ & $\dot{C}(10^{-5}yr^{-1})$ & $T_{2D}(H_{0}^{-1})$  &  $\bar{L} (ergs^{-1}) $ \\ 
    \hline
  0.001  & $0.14 \pm 0.06$  & $0.11\pm0.05$ & $3.9\pm1.7\cdot10^{39}$  \\
     0.01 & $ 0.38\pm0.17 $ & $ 0.39\pm0.18 $ & $1.1\pm0.5\cdot10^{40}$  \\
      0.05 & $0.8 \pm 0.4$ &$0.9\pm0.4$& $2.2\pm1.0\cdot10^{40}$  \\
   0.1 & $1.1 \pm 0.5$ & $1.4\pm 0.7$ & $3.0\pm1.4\cdot10^{40}$ \\ 
  0.25 & $1.6 \pm 0.8$  & $2.3\pm1.1$ & $4.5\pm2.2\cdot10^{40}$ \\
  0.5 & $2.2 \pm 1.0$ & $3.4\pm1.6$ & $6.2\pm3.0\cdot10^{40}$  \\
  1 & $3.0\pm1.4$ & $4.9\pm2.4$ & $8.4\pm4.1\cdot10^{40}$ \\
  2 & $4.0\pm 2.0$ &$7.3\pm3.6$ & $1.1\pm0.6\cdot10^{41}$ \\
  4 & $5.5\pm2.7$ &$11\pm5$ & $1.6\pm0.8\cdot10^{41}$ \\
  10 & $8.3\pm4.2$ &$18\pm9$ & $2.4\pm1.2\cdot10^{41}$  \\
    \hline
   \end{tabular}

\begin{tabular}{|c|}
  \cite{Ford2005} 
   \end{tabular}
  \begin{tabular}{|c|c|c|c|}
   $ \mbh (10^{6}{\msun})$ & $\dot{C}(10^{-5}yr^{-1})$ & $T_{2D}(H_{0}^{-1})$  &  $\bar{L} (ergs^{-1}) $ \\ 
     \hline
      0.001  & $0.40 \pm 0.17$  & $0.04\pm0.02$ & $1.1\pm0.5\cdot10^{40}$  \\
     0.01 & $ 0.90\pm0.40 $ & $ 0.16\pm0.07 $ & $2.6\pm1.2\cdot10^{40}$  \\
      0.05 & $1.6 \pm 0.7$ &$0.46\pm0.21$& $4.5\pm2.1\cdot10^{40}$  \\
   0.1 & $2.0 \pm 0.9$ & $0.72\pm 0.34$ & $5.8\pm2.7\cdot10^{40}$ \\ 
  0.25 & $2.8 \pm 1.3$  & $1.3\pm0.6$ & $8.0\pm3.8\cdot10^{40}$ \\
  0.5 & $3.6 \pm 1.7$ & $2.0\pm1.0$ & $1.0\pm0.5\cdot10^{41}$  \\
  1 & $4.6\pm2.2$ & $3.2\pm1.5$ & $1.3\pm0.6\cdot10^{41}$ \\
  2 & $5.8\pm 2.8$ &$5.0\pm2.5$ & $1.7\pm0.8\cdot10^{41}$ \\
  4 & $7.4\pm3.7$ &$7.9\pm3.9$ & $2.1\pm1.0\cdot10^{41}$ \\
  10 & $10.3\pm5.2$ &$14\pm7$ & $2.9\pm1.5\cdot10^{41}$  \\
  \hline

  \end{tabular}

\captionsetup{format=plain,labelsep=period,font={small}}
  \caption{The expected number of stellar disruption events $\dot{C}$ for solar like stars by supermassive black holes up to $\mbh \le 10^{7}\msun$. For comparison our numerical results
are extrapolated according to an older version of the $\mbh-\sigma$-relation
\citep{Ford2005} and the most recent one \citep{Schulze2009}. Within a factor of two they agree with each other.
 $T_{2D}$ is the time needed to double the initial mass of the black
hole in units of the Hubble time $H_{0}^{-1}$. Only one half of the stellar mass is assumed to become accreted by the black hole \citep{Rees1988}. Finally the time averaged mean luminosity $\bar{L}=0.5\epsilon \dot{C}\msun c^{2}$ of these black holes is calculated by assuming the efficiency parameter of matter to energy conversion to
be $\epsilon=0.1$. The motivation behind is to compare these energies with potentially detectable left overs of relativistic outflows which may become deposited into the surrounding medium after tidal disruption events
\citep{Crocker2011, Giannios2011, vanVelzen2011}. However the deposited energy strongly depends on the formation rate of relativistic jet outflows and may be significantly overestimated by us \citep{Bower2011}.
Nevertheless these deposited energies might be relevant for studies aiming to make a robust detection of dark matter annihilation signals in galactic bulges, dwarf galaxies or globular clusters hosting a central black hole.  
These results have relevance for galaxies with cuspy density profiles with slope parameters $\alpha\approx-1.5$ within the inner most few pc.}
\label{tablen=5}
\end{center}
\end{table}\\

The expected number of tidal disruption events in galactic nuclei containing black holes of $10^{6}$ to $10^{7}\msun$ inferred from the numerical integrations are in good agreement with recent optical
based surveys \citep{vanVelzen2009}. While their study yields the rate for tidal flares per galaxy to be $\dot{C} =3(^{+4}_{-2})\cdot10^{-5}\text{yr$^{-1}$}$, the results obtained from the present simulations give
$\dot{C} =3.0(\pm1.4)-8.3(\pm4.2)\cdot10^{-5}\text{yr$^{-1}$}$ for black holes in the mass range $10^{6}$ to $10^{7}\msun$. \\

The simulations also offer some clues about the growth of IMBHs and SMBHs in the lower mass range. We observe only a modest impact of the black hole mass on the capture rate. For a mass range over
four orders of magnitude, the capture rate increases only by a factor 25-60 depending on the used scaling relation. The relaxation driven growth of massive black holes by stellar disruptions is thus only important for 
IMBHs and SMBHs up to several $10^{5}\msun$. IMBHs should easily double their mass within a few Gyr in perfect agreement with earlier studies \citep{Baumgardt2004a}. Much more massive black holes must
have grown by different processes rather than the relaxation driven infall of stars, in good agreement with the findings of \cite{Yu2002} and gas accretion and feedback models
\citep{Silk1998, Fabian1999, Murray2005}\footnote{The growth of the very early population of SMBHs may also be dominated by stellar disruptions in isothermal cusps \citep{Zhao2002}. See the information in the text below.}.\\

Our findings exclude any relevance for establishing the $\mbh-\sigma$ relation from stellar disruptions in density profiles similar to those of the simulations. This is due to the relatively small capture rate and hence large doubling times ($T_{2D}>H_{0}^{-1}$) for black holes more massive than $10^{6}\msun$.
If for example the initial mass of a SMBH is strongly under-massive with respect to the $\mbh-\sigma$ relation, the feeding from tidal disruptions events alone might not be sufficient enough to bring it close to the
observed relation for galaxies at $z\approx0$. 
On the other hand if stellar disruptions dominate the growth of the least-massive black
holes there is no obvious reason why these black holes should follow the $\mbh-\sigma$ relation.
By now assuming the $\mbh-\sigma$ relation to be established for a primordial gas rich globular cluster (or galactic nucleus), which nowadays remains in isolation and without gas to drive new star formation,
the resulting IMBH (or SMBH, at least if it is not too massive) should nowadays be more massive than expected from the $\mbh-\sigma$ relation due to subsequent tidal
disruption events. It is very tempting to connect these results to the case of $\omega$-Centauri \citep{Noyola2010}. Tidal disruption events might therefore have implications for the search and existence of
IMBHs in globular clusters. Of course in order to proof its relevance for IMBHs, the use of the $\mbh-\sigma$ relation in the extrapolation formalism from numerical simulations to galactic nuclei (Appendix~\ref{AppendixB})
must be replaced by more direct observational data because the extrapolated values strongly depend on the validity of this scaling relation. Tidal disruption events complicate 
the understanding of the relevant processes which drive the evolution of galaxies and their central black holes. Especially as the impact of disruption events for the mass growth of black holes strongly 
depends on their initial mass. \\

In spite of this it might be interesting
to relate these findings to a recent study \cite{Kormendy2011} in which observational evidence for secular growth processes of black holes in disks and pseudobulges is found. The capture rate for 
rotation-supported models like rotating bulges or pseudo bulges should be enhanced compared to nonrotating models. These objects are expected to form from rotating bar instabilities \citep{KormendyPsedo}
and the relative velocities between two or more particles are generally lower. Therefore two-body relaxation processes would be even stronger. \\

However the overall picture of black hole growth across cosmic times by tidal disruptions might be complicated even more due to the dynamical evolution of the density profile and a variable fraction of
the initial stellar mass which finally becomes accreted by the black hole. Our conclusions regarding the growth history of IMBHs and SMBHs events should 
only hold for density profiles resembling those of our simulations and by assuming that a fraction of one half (or more) of the initial stellar mass becomes accreted by the black hole \citep{Rees1988}. 
In fact some effects can considerably reduce this fraction and complicate the efforts to estimate the significance of tidal disruption events for the overall growth history of black holes. Recent hydrodynamical
simulations suggest that for loss cone stars on nearly parabolic orbits, most of the stellar matter is ejected within the first orbit and then later on due to powerful shocks which may be energetic enough to ignite thermonuclear reactions
unbinding large amounts of stellar mass \citep{Brassart2008, Guillochon2009}. Secondly and especially relevant for
black holes in the lower mass range, accretion luminosities far in excess of the Eddington limit \citep{Strubbe2011} may blow away most of the remaining gas. In the end
the growth of these black holes due to tidal disruption events may be insignificant even for very large capture rates of several events per $10^{6}$\unit{$\text{yr}$}. \\

\section{Critical discussion and outlook for future work}\label{5.3}
To the best of our knowledge this study reports for the first time the expected tidal disruption rate of stars by SMBHs up to $10^7\msun$ obtained by direct N-body integrations. N-body computations offer
a large amount of advantages over analytical studies. They can handle several physical effects simultaneously while most analytical studies are forced to simplify at least some of the dynamics.
On the other hand direct N-body integrations aiming to infer astrophysically relevant numbers of stellar disruption events are confronted by their own limitations and difficulties.
In this section we will critically review limitations of our own simulations as well as improvements and ideas left for future work.
 \begin{enumerate}
  \item In Table~\ref{tablen=5} we calculate among other values the required timescale $T_{2D}$ for doubling the mass of a black hole of given initial mass. This timescales is computed from the total number of captures 
        averaged over 100 N-body time units (see Table~\ref{tablen=4}). We recommend the reader to regard the doubling time $T_{2D}$ only as some reference guide. When expressed in physical
        time, our simulations last only a fraction of one $H^{-1}_{0}$
        (between several $10^7$ and one $10^9$ years) and may not represent much longer time episodes. Moreover we assumed one half of the disrupted star to be accreted by the black hole. There exist two effects that
        can reduce the amount of stellar matter which finally becomes swallowed by the black hole. First, if the
        tidal stripping occurs from a nearly parabolic orbit, hydrodynamical simulations suggest one half of its mass to be lost within its first path \citep{Guillochon2009} and large quantities of the remaining mass to be blown away by shocks and thermonuclear reactions later on \citep{Brassart2008}.
        Second, very small black holes might temporarily generate luminosities far in excess of the Eddington limit \citep{Strubbe2011} and most of the remaining 
        matter may finally be blown away instead of being swallowed by the black hole. This would invalidate our conclusions regarding the growth history of small black holes where we assumed one half of the stellar 
        mass to be accreted.
        Nevertheless the inferred capture rate should be valid for all galaxies or stellar clusters with density profiles
        comparable to our simulated ones. 
  \item With current generations of GPUs it is unthinkable to simulate galaxy models with realistic numbers of stars with direct N-body integration methods. The only way to obtain stellar disruption
        rates for SMBHs in the centers of galaxies is to simulate as many models as possible to infer all relevant N-dependent systematics affecting this rate. Afterwards the results can be extrapolated. However it is 
        important not to do this for only one given black hole capture radius but for many black hole configurations. Therefore all these simulations must be repeated for several capture
        radii $r_{\text{cap}}^{\text{sim}}$ in order to
        extrapolate them according to the formalism in \S~\ref{5.1} to the black hole of interest. We calculate the capture rate for black hole masses in the range $\unit[10^{3-7}]{\msun}$. Due to the highly
        nonlinear Eq.~\ref{F57a} we had to extrapolate parameter $r_{\text{cap}}^{\text{sim}}$ from Table~\ref{tablen=4} down to $r_{\text{cap}}^{\text{sim}}\approx0.07\cdot10^{-7}$. The usage of three different black hole capture radii
        is thus the minimal requirement to obtain useful values under the assumption that the parameter $a(r_{\text{cap}}^{\text{sim}})$ from Eq.~\ref{FP} follows a power law distribution\footnote{$a(r_{\text{cap}}^{\text{sim}})$ is specified in Eq.~\ref{Fa}.} with positive parameters and no offset.\\

        There is no question that future studies must redo these simulations for different capture radii to constrain $a(r_{\text{cap}}^{\text{sim}})$ even more precisely. However this is a very time consuming task.
        The complete set of our S\'{e}rsic $n=4$ simulations took more than seven months to compute on
        five modern GPUs. Despite the large amount of needed computing power, direct integration methods like NBODY6 may exceed their limitations when the capture radius falls significantly below $10^{-7}$ in N-body units,
        especially if the mass of the black hole particle is
        of the order of one percent or more of the total mass\footnote{Private communication with Sverre Aarseth.}. In addition to that the statistics may worsen (due to a limited number of capture events) and must be balanced
        by even more simulations.\\

        In our computations no severe $r_{\text{cap}}^{\text{sim}}$ dependence of the parameter $b$ is evident, in accordance with theoretical considerations
        (Eq.~\ref{F18} \& \ref{F20})\footnote{According to the theory of angular momentum diffusion, parameter $b$ only depend on the slope parameter of the density profile.}. Therefore we assumed it to be constant.
        However we cannot exclude per se any deviation at very small capture radii. A systematic decrease in the parameter $b$ for even smaller values of $r_{\text{cap}}^{\text{sim}}$ would only reduce the tidal disruption events
        of the more massive SMBHs in our sample.
        We plan to tackle this problem as well as to constrain the parameters
        $a(r_{\text{cap}}^{\text{sim}})$ and $b$ even more precisely in the future.    
  \item In this study effects from General Relativity are neglected. The relevant tidal disruption radius of a SMBH for solar 
        like stars is several times larger than its Schwarzschild radius and relativistic effects should become strongly suppressed for radii $r>>r_{s}$. This makes our assumption of neglecting GR credible.
        Nevertheless a fully relativistic treatment of a black hole potential yields a deeper gravitational potential than a purely Newtonian one,
        thus being more attractive for compact bodies like stars to be captured by the SMBH. On the other hand particle scattering by a relativistic potential may result in stronger deflection, perhaps
        powerful enough to reject some stars from the immediate vicinity of the black hole thereby decreasing the capture rate. The next generation of N-body integrators is expected to be sophisticated enough to address these aspects \citep{Aarseth2007}. \\

        Rotating black hole spacetimes should be considered, too. 
        The cross section of a realistic black hole strongly depends on its mass and spin parameter $j=\frac{\jmbh}{\mbh}$, where $\jmbh$ is the angular momentum of the black hole. The likelihood
        for a particle to become swallowed by the black hole depends on the spin parameter, its trajectory and angular momentum.
        Particles are more likely captured if they counter rotate the black hole because in this direction the effective capture radius is enlarged. It is not unreasonable to conclude that a rotating black hole
        embedded inside a nonrotating spherical distribution of stars will lose some of its angular momentum. On the one hand $j$ decreases when $\mbh$ becomes larger,
        on the other hand counter rotating particles are more likely
        captured. This would lower $\jmbh$ and thus the spin parameter $j$. If tidal disruption events really contribute a significant amount of mass to a specific population of black holes, it should also affect
        their spin values in a way which might deviate from the predictions of gas accretion models. 
  \item A crucial quantity for extrapolating our numerical results to astrophysical systems is the black hole radius of influence $r_{H}$. For its evaluation we use the kinematic determination (Appendix~\ref{AppendixB}).
        We observe this radius to be roughly five to six times smaller than the dynamical radius $r_{g}$. This is the radius at which the mass in stars/particles equals the mass of the black hole. If interested readers plan to
        rescale our models by replacing the $\mbh-\sigma$ relation by directly measured data of $r_{H}$ for some galaxies, it is very important that they also use the same influence radii as the ones used in our simulations and not the 
        dynamical radii.     
  \item The capture rate from our numerical results should not be applied to SMBHs above $10^{7}\msun$. The refill of the loss cone takes a timespan of the order $T_{\text{refill}}\approx \theta_{\text{lc}}^{2}T_{\text{rel}}$.
        The refill of the loss cone is much faster in N-body integrations than in nature, since the potentials are more cuspy and relaxation times are shorter than in reality. Therefore our simulations have only relevance for
        galactic nuclei where the loss cone refilling times are much shorter than $H_{0}^{-1}$. In Appendix~\ref{AppendixC} we show that $T_{\text{refill}}<<H_{0}^{-1}$ 
        for a black hole with a mass of $10^{7}\msun$. This becomes also evident from Eq.~\ref{F18} \& \ref{F20}.
        For black holes significantly more massive than $10^{7}\msun$, i.e with very large particle numbers and very smooth potentials, the critical radius becomes much larger than the influence radius of the black hole and Eq.~\ref{F18}
        has to be replaced by Eq.~\ref{F20}. The latter one predicts a different behavior for $\dot{C}(N)$ such that the numerically found capture rate should not be extrapolated to black holes in excess of $10^{7}\msun$.   
        By inserting the relevant values from or computational findings to the systems of interest, Eq.~\ref{F18} predicts the critical radius not to exceed the influence radius for black holes less massive than $10^{7}\msun$, thus showing our 
        simulations to be governed by processes $r_{\text{crit}}<r_{H}$. 
  \item One could even criticize the black hole mass $\mbh(t=0)=0.01$ used for our numerical computations to be too high as the black hole mass fraction in realistic galaxies is a factor of a few smaller \citep{Magorrian1998}. Nevertheless 
        most of the relevant dynamics happens at distances of the order of the influence radius $r_{H}$ whereas we use the radius of influence for the extrapolation to realistic galaxies. The choice of $\mbh(t=0)=0.01$ is therefore not 
        expected to change the capture rate significantly. 
        In this context the usage of different capture radii instead of different initial masses $\mbh(t=0)$ for the extrapolation to the wide set of astrophysical SMBHs should be justified, too.
        The strict relation between mass and capture radius of a black hole (Eq.~\ref{F61}) enables variation of the latter one while keeping the former one constant in the scale-free N-body simulations.
        The great advantage of this strategy is given in equal black hole influence radii, crossing times, cusp formation timescales etc. simplifying the extrapolation formalism considerably.     
        The same holds true for the overall S\'{e}rsic $n=4$ profiles. Not every outer bulge component or elliptical galaxy profile resembles that of a S\'{e}rsic $n=4$ i.e. 
        de Vaucouleurs profile. Mostly relevant for the direct number of capture events is the density profile close to $r_{H}$. For relaxation times smaller than one $H_{0}^{-1}$ the formation of a cusp (up to $\alpha=-1.75$) is expected.
        Such a gradual change of the density profile is also found in the numerical simulations.
        Hence our simulations cover a large space of isotropic, nonrotating density profiles for black hole masses up to $\unit[10^{7}]{\msun}$.  
  \item We only treat single-mass systems while galactic cores are known to be multiple-mass systems featuring additional processes like mass segregation, star formation, binary evolution,
        torques from anisotropic matter distributions, resonances etc. Stellar remnants like neutron stars would not be disrupted outside the event horizon and could probe much deeper potentials than solar like stars, thus complicating the gravitational dynamics and making relativistic correction terms
        inescapable. They would also disappear without any visible counterpart when finally captured.  
  \item Finally our numerical simulations should only be regarded as a first (very) limited approach to a systematical scan of capture rates in galaxies. It would be important to extent these studies by simulating the same models
        for even smaller capture radii $r_{\text{text}}$ and longer timescales in order to reduce the need of extrapolation.
        It would be important to take into account rotating and triaxial stellar density profiles around the SMBH and to decrease the still rather large uncertainties. 
        Direct N-body simulations of isothermal $\rho(r)=\frac{\sigma^{2}}{2\pi G r^{2}}$ spheres, which might represent the initial phases of elliptical galaxies and bulges best, should be performed as well.
        \cite{Zhao2002} found evidence for strong black hole growth in isothermal cusps. 
        By assuming $r_{\text{cap}}\propto r_{s}=\frac{2G\mbh}{c^{2}}$, rewriting Eq.~\ref{F17} to $\dot{C}(r)\propto \rho(r)r^{2}\sigma \theta_{lc}^{2}$ by using $r_{H}=\frac{G\mbh}{\sigma^{2}}$ and
        $\theta_{lc}^{2}=\frac{2r_{\text{cap}}r_{H}}{r^{2}}$ for very massive SMBHs, one obtains
        $\dot{C}(r)\propto \frac{\sigma^{5}}{Gc^{2}}\cdot \left(\frac{r_{H}}{r} \right)^{2}$. Under the assumption that the capture rate is dominated by stars from $r_{H}$, the $r$ dependence cancels out and the final mass
        of the black hole is $\mbh(t_{f})=\int^{t=t_{f}}_{t=0}\dot{C}(r=r_{H})\text{d}t\approx 10^{8}\msun \cdot \left(\frac{\sigma}{200\text{kms}^{-1}}\right)^{5}\left(\frac{t_{f}}{H_{0}^{-1}} \right)$. This relation is indeed in
        very close agreement to the observed $\mbh-\sigma$ relation \citep{Zhao2002}. Therefore stellar captures might contribute significantly to the growth of SMBHs in the past, especially if the loss cone refill
        is enhanced by mergers and/or triaxial stellar distributions.\\         
 \end{enumerate}

Despite some of the details stated above, the here reported simulations represent (the first) systematic estimate for the capture rate by SMBHs of stars in galaxies with cuspy inner density profiles.
This work should be followed up
by simulating different capture radii $r_{\text{cap}}^{\text{sim}}$ as well as density profiles, taking relativistic correction terms into account, and by trying to find ways to infer the numbers of
disruption/capture events for SMBHs with mass $>\unit[10^{7}]{\msun}$.
 
\section{Conclusion}\label{7}
We performed direct N-body simulations to obtain the number of disruption events of stars by SMBHs which are presumed to exist in the centers of most galaxies. A modified NBODY6 code was used.
All computations were processed by several GPUs over several months integration time. The initial density profiles of the models were chosen to follow nonrotating isotropic S\'{e}rsic $n=4$ profiles.
We calculated numerous models with
different particle numbers but otherwise equal physical parameters in order to ensure good statistics. This is required because all systematic effects depending on the total number of particles must be specified
in order to extrapolate the simulations to realistic galaxies
by using the formalism presented in \S~\ref{5.1}. The rates at which stars are captured are found to be nearly independent of the mass of the black hole. Thus only the growth over cosmic times of IMBHs
and of the least massive SMBHs may be dominated by stellar disruptions.
The expected tidal disruption rate is a few events every $10^{5}$ years per galaxy for black holes in the mass range up to $\unit[10^{7}]{\msun}$. The feeding by stars from density profiles similar to the ones computed here
bears no implications for establishing scaling relations
between very massive black holes and their host galaxies. This is in agreement with conventional gas accretion/feedback models.
On the other hand the growth history of the least massive black holes might be governed by more than one feeding mode (gas and star accretion). This might have implications for the search and existence of potential
IMBHs in globular clusters and minor galaxies. Assuming these scaling relations (e.g. the $\mbh-\sigma$ relation) to be established shortly after their primordial gas rich phase billions of years ago, the nuclear
black holes would continue their growth by the subsequent disruption of stars. Depending on the initial conditions,
the black hole masses could nowadays lie well above the predicted values of the $\mbh-\sigma$ relation as long as the globular cluster remains in isolation\footnote{The relevant velocity dispersion $\sigma$
should therefore not increase. }. On the other hand the continuous monitoring and search for tidal disruption events in globular clusters (e.g. in the Virgo Cluster) should constrain the fraction of those clusters
hosting a central IMBH. By assuming 25000 globular clusters with a central black hole in the mass range $\mbh=10^{3}-10^{4}\msun$ in the Virgo Cluster of galaxies, there should be one disruption event every $10-25$ years.
Finally the performed computations indicate that the growth history of IMBHs and low mass SMBHs is diverse and not only governed by one process, i.e gas accretion.
However it needs to be pointed out that there exist effects which might reduce the fraction of stellar matter which finally becomes accreted by the black hole.
We assumed one half of a captured star's mass to be swallowed \citep{Rees1988}, whereas a smaller fraction would result in even slower growth rates. Thus our conclusions regarding the growth history may change if small black holes gather
only tiny fractions of the total initial stellar mass. Future studies can use the reported capture rate $\dot{C}(\mbh)$ to deduce more realistic growth rates $\dot{M}(\mbh)$ by taking more appropriate values for the fraction of
accreted matter into account.
It would also be interesting to extend these studies to the most-massive
black holes as well as constraining the capture rate for different profiles. 

\section*{Acknowledgments}
We warmly thank Ole Marggraf and Fabian L\"ughausen for their readiness to assist us in some technical problems as well as Sverre Aarseth, Jan Pflamm-Altenburg and Sambaran Banerjee for inspiring discussions about black holes and numerical 
integrators. Special thanks are devoted for Ingo Thies and Matthias Kruckow. The work of this paper was supported by the German Research Foundation
(DFG) through grants BA 2886/4-2 within the priority programme 1177 ``Witnesses of Cosmic History: Formation and Evolution of Black Holes, Galaxies and Their Environment''.
H.B. acknowledges support from the Australian Research Council through Future Fellowship grant FT0991052.    
\bibliographystyle{mn2e}
%\bibliography{Bibtex}

\begin{appendix}
 
\section{The tidal disruption/capture radius}\label{AppendixA} 
The disruption radius at which a star is torn apart by tidal forces such that roughly one half of its matter will become accreted by the black hole, is a function
of the mass and spin of the black hole as well as the trajectory, internal structure, size and mass
of the star. 
A star is disrupted outside the event horizon if the black hole mass is smaller than a certain limit \citep{Lai1994, Binney2008}. 
For typical solar-type stars with masses $\mstar \approx 1\msun$ and radii $\rstar\approx1\rsun$ the mass of the black hole must be smaller than
$\mbh \leq 10^{8}\msun$ to disrupt the star before reaching the event horizon. A strongly spinning black hole dramatically alters the situation \citep{Ivanov2006}.
Sufficiently massive black holes swallow stars as a whole. The General Theory of Relativity predicts the radius where a star is doomed to enter a very massive 
black hole to be larger than the actual Schwarzschild-radius $r_{s}$ \citep{Novikov1989}. Stars on initial Keplarian orbits coming from infinity with pericentre
distances $q\leq4r_{s}$ will be
swallowed by the black hole as long as $\big(\frac{v_{\infty}}{c} \big)^{2}\ll 1$. The particles in the N-body simulations do not come from infinity but their
speed at the apocentre distance is much lower than the corresponding speed of light and the capture radius of $r_{\text{cap}}=4r_{s}$ seems to be the most natural and
best approximation for the behaviour of a realistic (extremely massive $\geq 10^{8}\msun$) black hole. This approximation also holds for
the bound particles around the black hole which are most likely swallowed.
The ratio $(\frac{v_{\text{apo}}}{v_{\text{peri}}})^{2}$ for apocentre distances of $10^{-4}-10^{-2}$ is always much smaller than one.
For our purposes the capture radius can finally be defined as:
\begin{equation}\label{F61}
 r_{\text{cap}}= \left\{ \begin{array}{l@{\quad:\quad}l}
 g \rstar \left(\frac{\mbh}{\mstar}\right)^{\frac{1}{3}}  &   \mbh \lesssim  10^{8}\msun \\
 \frac{8G\mbh}{c^{2}}  & \mbh \gtrsim 10^{8}\msun
 \end{array}  \right. \
\end{equation}
The parameter $g$, which is of the order of one, depends on many parameters and can be taken from \cite{Kochanek1992, Lai1994, Ivanov2006}.

\section{Extrapolation}\label{AppendixB} 

In the following part we give a more detailed description of the formalism by which the here obtained capture rates (Table~\ref{tablen=4}) can be scaled up to realistic bulges of galaxies or elliptical galaxies.
\begin{enumerate}
 \item From the relation 
\begin{equation}\label{F59}   
\frac{r_{\text{cap}}}{r_{H}}\Big|_{\text{sim}}=\frac{r_{\text{cap}}}{r_{H}}\Big|_{\text{astro}}
 \end{equation}
the required capture radius $r_{\text{cap}}^{\text{sim}}$ for a black hole of mass $\mbh$ must be obtained by using astronomical observations of individual galaxies or by  
making use of the $\mbh-\sigma$ relation from \cite{Schulze2009}.
If in the near future much larger samples of measured SMBH masses allow for more accurate values, it will be no problem to implement them into this formalism.
By combining Eq.~\ref{F59} with the disruption radius $r_{\text{cap}}=g \rstar \big(\frac{\mbh}{\mstar}\big)^{\frac{1}{3}}$ and the expression for the radius 
of influence $r_{H}\approx 13.1\left(\frac{\mbh}{M_{8}}\right)^{0.54}[\text{pc}]$ which is derived from the $\mbh-\sigma$ scaling relation, $r_{\text{cap}}^{\text{sim}}$ follows:
\begin{equation}\label{F59a} 
 r_{\text{cap}}^{\text{sim}} \approx 4g\cdot 10^{-9} \left(\frac{\mbh}{M_{8}}\right)^{-0.2067}.
\end{equation}
It specifies the required capture radius in the scale-free N-body integrations for the astrophysical black hole of interest.
Afterwards the function $a(r_{\text{cap}}^{\text{sim}})$ must be evaluated from the values in Table~\ref{tablen=4}:
\begin{equation}\label{Fa} 
a(r_{\text{cap}}^{\text{sim}})=0.023(\pm0.006)\left(r_{\text{cap}}^{\text{sim}}\right)^{0.363(\pm0.020)}
\end{equation}
yields a reasonable approximation\footnote{$Q=0.89$ without rescaling 
$\chi_{\mu}=1$. Afterwards the uncertainties are taken directly from the covariance matrix. Renormalization induces the errors to be uncorrelated to each other.} for the extrapolation of the parameter $a$ from Eq.~\ref{FP} to any desired $r_{\text{cap}}^{\text{sim}}$.
For the purposes of this paper the slope parameter $b=0.83$ is assumed to be independent of $r_{\text{cap}}^{\text{sim}}$.  
As already mentioned in \S~\ref{5.1} the parameter $g$ accounts for the stellar model and mass of the black hole.
It is of the order of one \citep{Kochanek1992, Lai1994, Ivanov2006}. For simplicity we use $g=1$ which is a reasonable assumption for nonrotating black holes less massive than $\mbh=10^{7}\msun$ and 
solar mass stars. Eq.~\ref{F59a} assumes all stars to be disrupted before entering the horizon.
\item The dynamical timescale $t_{\text{sim}}$ of the N-body particles inside the sphere of influence $r_{H}$ has to be calculated according to $t_{\text{sim}}=\frac{2r_{H}}{\sigma(r=r_{H})}\approx 0.008$.
It is used as a reference for timing issues when compared to the relevant astrophysical timescales $t$. 
To ease the extrapolation of the numerical results to astrophysical systems, we compute the time averaged influence radius $r_{H}$.
Representative for all models we calculate $r_{H}$ and $t_{\text{sim}}$ from the \unit[$25$]{$\text{k}$}, \unit[$50$]{$\text{k}$}, \unit[$75$]{$\text{k}$}, \unit[$150$]{$\text{k}$} and \unit[$250$]{$\text{k}$} models.
For the calculation of the radius of influence we bin the particles in cylindrical shells of thickness $\Delta r=0.001$ and measure for each configuration
the one dimensional velocity dispersion (line of sight velocity) $\sigma_{i}^{2}=\frac{\sum_{i} v_{i,z}^{2}}{N_{i}}$ in order to obtain $\sigma(r)_{\text{sim}}^{2}$. Here $N_{i}$ is the number of particles within each configuration.
We choose the line of sight axis to
be parallel to the z-axis.
Afterwards $\sigma_{\text{bh,i}}^{2}=\frac{\mbh(t)}{3N_{i}}\cdot \left(\sum_{i=1}^{N_{i}} \frac{1}{r_{i}} \right)$ is calculated for each cylindrical shell to obtain $\sigma(r)_{\text{bh}}^{2}$, here $r_{i}=\sqrt{x_{i}^{2}+y_{i}^{2}+z_{i}^{2}}$.
The factor $3$ in the denominator is used for the normalization to the relevant line of sight velocity inside the isotropic distribution. The radius of influence $r_{H}$ is then calculated to be the radius at which
$\frac{\sigma(r)^{2}_{\text{sim}}}{\sigma(r)^{2}_{\text{bh}}}=2$.
We note that in N-body units $G=1$.       
The position of the black hole is used as the reference center and the mass gain of the black hole is taken into account. For the time averaged influence radius and velocity dispersion we obtain $r_{h}\approx0.005$ and 
$\sigma(r = r_{H})\approx 1.26$. The black hole influence radius is $5-6$ times smaller than the dynamical radius.  
\item Subsequently the astrophysical dynamical timescale $t_{\text{cr}}(r_{H})=\frac{2r_{H}}{\sigma}\big|_{\text{astro}}$ of the matter distribution within the influence radius
of the astrophysical galaxy must be computed for the black hole of given mass by using $r_{H}\approx 13.1\left(\frac{\mbh}{M_{8}}\right)^{0.54}[\text{pc}]$ and $\sigma \approx 200\Big(\frac{M_{8}}{1.5135}\Big)^{0.23}[\text{kms}^{-1}]$ from \cite{Schulze2009}. 
\item The number of stars $N$ in the astrophysical galaxy must be specified. For simplicity we assume all stars to have the same mass $\langle \mstar \rangle = 1\msun$.
A coarse estimate for the number of stars can be computed by:
\begin{equation}\label{F58}
N= \frac{100\mbh}{\langle \mstar \rangle}.
 \end{equation}
The choice of $\langle \mstar \rangle = 1\msun$ depends on the stellar mass function and seems to be a reasonable assumption for galactic nuclei where 
mass segregation is important \citep{Freitag2006, Kroupa2001, Lockmann2010}.
The factor 100 accounts for the fraction of bulge mass to black hole mass in accordance with our simulations.
\item Finally the disruption rate of stars by massive black holes can be evaluated. In a first step the numerically inferred number of captures $\dot{C}(N)$ per N-body time unit (Table~\ref{tablen=4}) must be normalized
to the relevant crossing time $t_{\text{sim}}=0.008$ (in N-body time units) at the influence radius of the black hole.
This dimensionless number must afterwards be synchronized with the relevant timescale $t_{\text{cr}}(r_{H})$ of the
astrophysical galaxy. Consequently $\dot{C}(N)\cdot t_{\text{sim}}$ has to be divided by $t_{\text{cr}}(r_{H})$ in order to obtain the number of disrupted stars within the desired physical time unit (e.g yr, Myr) for the black hole of interest:  
\begin{equation}\label{F56}
 \dot{C}_{\text{astro}}=\frac{0.008 \cdot a(r_{\text{cap}}^{\text{sim}})N^{b}}{t_{\text{cr}}(r_{H})}.
\end{equation}
\end{enumerate}
Our extrapolation formalism strongly depends on the $\mbh-\sigma$ relation. More accurate and numerous black hole measurements will improve this relation in the future.
Moreover we only treat errors from our simulations and neglected the intrinsic scatter of the $\mbh-\sigma$ relation for simplicity.

\section{Loss cone problems}\label{AppendixC} 
Direct N-body integrations are limited by a maximal computable number of particles which is orders of magnitudes lower compared to particle numbers in the nuclei of astrophysical galaxies. The extrapolation to such
astrophysical settings is thus only possible if the relevant physics do not change in between. Loss cone problems (tidal capturing and/or shrinking binary black holes) require special care \citep{Gualandris2007}.
Here we will show that the inequality 
\begin{equation}\label{F67}
 T_{\text{refill}}=\theta^{2}T_{\text{rel}}<<H_{0}^{-1}
\end{equation}
is fulfilled up to SMBHs of order $10^{7}\msun$ and hence our result, $\dot{C}\propto N^{0.83}$, should yield realistic values when extrapolated to such black holes. For black holes much more massive, the situation 
might change. By assuming the radius $r$ at which particles can enter loss cone orbits without being scattered away through interactions with other stars to
be $r_{\text{crit}}\approx r_{H}\big|_{\mbh=10^{7}\msun}$, the loss cone angle $\theta$ can be evaluated from Eq.~\ref{F13}. For the constant of proportionality $f$ we use $f=2$ in accordance with \cite{Rees1976}.
%The angular momentum is $L=r\sin({\theta})v \mstar \approx r_{H}v_{T}\mstar$ whereas the 
%pericentre distance $d_{\text{min}}$ must be equal or even smaller than $r_{\text{cap}}$:
%\begin{equation}
% r_{\text{cap}}\ge d_{\text{min}}=a\cdot(1-\epsilon).
%\end{equation}
%The eccentricity parameter $\epsilon=\sqrt{1-\frac{b^{2}}{a^{2}}}$ is defined over the semi major $a=-G\frac{\mstar \mbh}{2E}$ and and minor axis $b=\frac{L}{\sqrt{-2\mstar E}}$,
%here $E=\frac{1}{2} \mstar v^{2}-G\frac{\mstar \mbh}{r} \approx -\frac{1}{2}\mstar v^{2}$ is the total energy of the strongly bound star. Subsequently the loss cone angle $\theta$ can be derived:
%\begin{equation}
% \theta \le \arcsin \left({\frac{\sqrt{-2\mstar E\cdot(-a^{2}(1-\frac{r_{\text{cap}}}{a})^{2}+a^{2})}}{\mstar v r}}\right).
%\end{equation}
By assuming $r\approx r_{H}$, $\mstar\approx 1\msun$, the relaxation time to be $T_{\text{rel}}\approx H_{0}^{-1}$ \citep{Freitag2008} and estimating all other relevant parameters from the $\mbh-\sigma$
relation \citep{Ford2005}, one obtains the desired result $T_{\text{refill}}\approx 4\cdot 10^{-6}\cdot H_{0}^{-1}<<H_{0}^{-1}$.
Even though our assumptions are idealistic and not every star fulfills its plunge into the 
black hole from the critical radius $r_{\text{crit}}$, it underlines the extrapolation from our numerical simulations to realistic cores of galaxies with central black holes up to $10^{7}\msun$ to be credible. 
\begin{figure}
\begin{center}
\includegraphics[width=8.5cm]{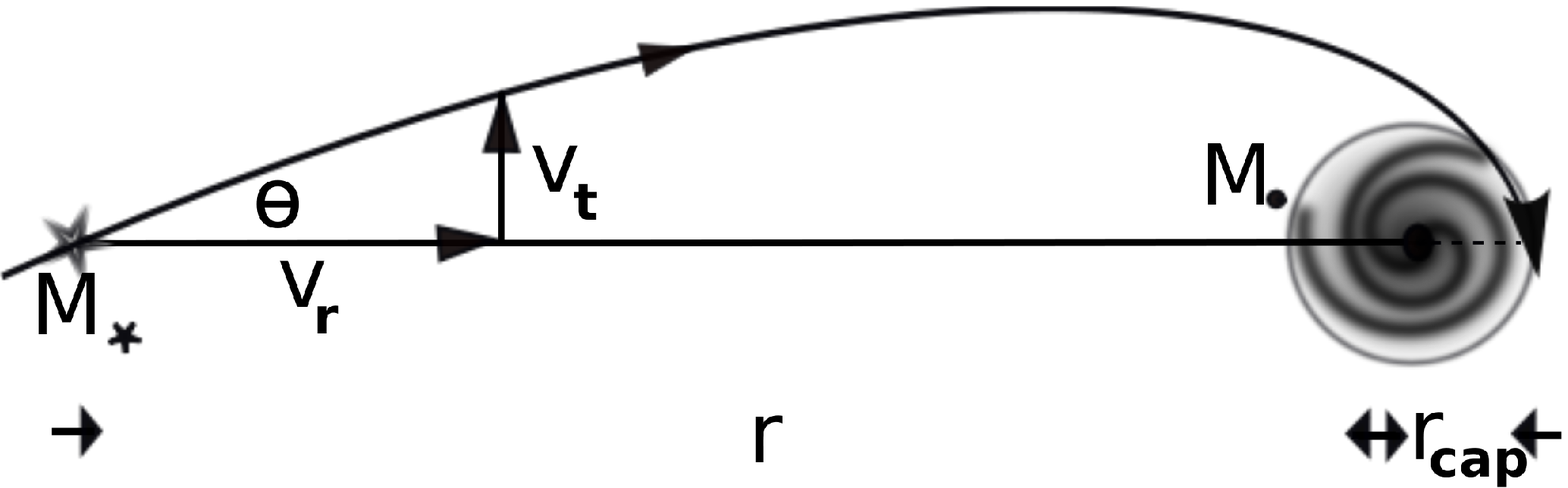}
\end{center}
\captionsetup{format=plain,labelsep=period,font={small}}
\caption{Sketch of a typical loss cone problem.}  
\label{losscone.eps.eps}
\end{figure}

\end{appendix}

%\bibliographystyle{mn2e}
%\bibliography{ithies_refs}

\label{lastpage}

\end{document}